\documentclass{appolb}
\usepackage{epsfig}

\begin{document}

\topmargin = 0cm

{
\begin{titlepage}

\begin{center}
\mbox{ } 

\vspace*{-3cm}

\vspace*{-2cm}


\end{center}
\vskip 0.5cm
\begin{flushright}
\Large
\mbox{\hspace{10.2cm} hep-ph/9911445} \\
\mbox{\hspace{10.2cm} IEKP-KA/99-21} \\
\mbox{\hspace{11.5cm} Nov. 1999}
\end{flushright}
\Large
\begin{center}
\vskip 2cm
{\huge\bf
Search for \\
New Particles

\vspace*{2mm}
at LEP}
\vskip 1cm
{\LARGE\bf 
Andr\'e Sopczak}

{\Large
\vspace*{2cm}
Karlsruhe University}
\end{center}

\vskip 1.8cm
\centerline{\Large \bf Abstract}

\vspace*{2.5cm}
\hspace*{-3cm}
\begin{picture}(0.001,0.001)(0,0)
\put(,0){
\begin{minipage}{16cm}
\Large
\renewcommand{\baselinestretch} {1.2}
The latest preliminary results of the searches 
for Higgs bosons and Supersymmetric particles 
at LEP are reviewed. The results include
the data-taking in 1999 up to center-of-mass energies 
of 196 GeV.
The combination of the results from the four 
LEP experiments leads to a significant increase
of the detection sensitivity. No indication
of a signal has been observed.
In the Standard Model (SM)
a lower limit of 102.6 GeV on the
mass of the Higgs boson is set at 95\% CL.
In extended models, stringent limits are also set on the
HZZ coupling. Interpretations in the 
Minimal extension of the Supersymmetric Standard
Model (MSSM) are given and the importance of
general MSSM parameter scans is emphasized.
In general scans, the limit on the mass of the lightest 
scalar Higgs boson is about 7 GeV lower in comparison with 
benchmark results.
The data also constrains charged Higgs bosons of
a general two-doublet model and Supersymmetric
partners of the SM particles.
\renewcommand{\baselinestretch} {1.}

\normalsize 
\vspace{1.5cm}
\begin{center}
{\large \em
Talk at the XXIII School of Theoretical Physics,
Ustron, Poland, Sep. 1999, \\ 
to be published in the proceedings.
\vspace*{-7cm}
}
\end{center}
\end{minipage}
}
\end{picture}
\vfill

\end{titlepage}

\newpage
\thispagestyle{empty}
\mbox{ }
\newpage
\setcounter{page}{1}
}

\title{Search for New Particles at LEP
\thanks{Presented at the XXIII School of Theoretical Physics,
        Ustron, Poland, Sep. 1999.}%
}
\author{Andr\'e Sopczak
\address{Karlsruhe University}
}
\maketitle
\begin{abstract}
The latest preliminary results of the searches 
for Higgs bosons and Supersymmetric particles 
at LEP are reviewed. The results include
the data-taking in 1999 up to center-of-mass energies 
of 196 GeV.
The combination of the results from the four 
LEP experiments leads to a significant increase
of the detection sensitivity. No indication
of a signal has been observed.
In the Standard Model (SM)
a lower limit of 102.6 GeV on the
mass of the Higgs boson is set at 95\% CL.
In extended models, stringent limits are also set on the
HZZ coupling. Interpretations in the 
Minimal extension of the Supersymmetric Standard
Model (MSSM) are given and the importance of
general MSSM parameter scans is emphasized.
In general scans, the limit on the mass of the lightest 
scalar Higgs boson is about 7 GeV lower in comparison with 
benchmark results.
The data also constrains charged Higgs bosons of
a general two-doublet model and Supersymmetric
partners of the SM particles.

\end{abstract}

\section{Introduction}

\begin{figure}[htbp]
\caption{\label{fig:lumi} Delivered integrated 
                          luminosity per LEP experiment.}
\vspace*{-2mm}
\begin{center}
\includegraphics[width=\textwidth,%
bbllx=218pt,bblly=307pt,bburx=396pt,bbury=452pt,clip=]
{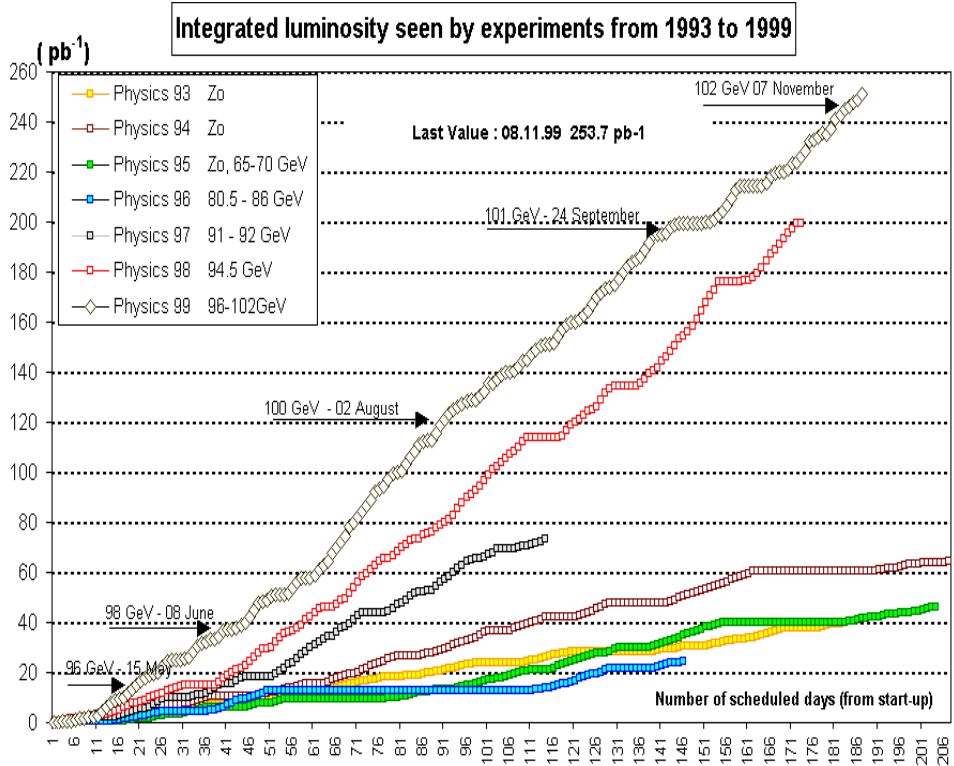}
\end{center}
\vspace*{-0.6cm}
\end{figure}

The data-taking at LEP has been very successful,
leading to continuously increasing center-of-mass energy and luminosity
as shown in Fig.~\ref{fig:lumi}. 
In August 1999, exactly 10 years after 
the first recording of a Z boson at LEP, 
a center-of-mass energy of 200 GeV was reached. 
The LEP accelerator will continue operation 
at a center-of-mass energy around 205 GeV in 2000.

The results of this report are mainly based on the combination
of the data from the four LEP experiments 
ALEPH, DELPHI, L3 and OPAL~\cite{higgswg}.
In the search for the Standard Model Higgs boson, all four
LEP experiments have contributed. Three experiments have
released their 1999 data for combination in the MSSM search, and
two for the charged Higgs boson search, 
leading to the luminosities shown in Table~\ref{tab:lumi}.

\begin{table}
\caption{\label{tab:lumi} 
Data used for the combination of the 
Higgs boson searches from the four LEP experiments.}
\vspace*{-4mm}
\begin{center}
\begin{tabular}{c|c|c|c|c}
    &      & SM Higgs & MSSM Higgs & Charged Higgs  \\
$\sqrt{s}$ (GeV)& Year & ${\cal L}$ (pb$^{-1}$) 
& ${\cal L}$ (pb$^{-1}$) & ${\cal L}$ (pb$^{-1}$) \\ \hline
189 & 1998 &  683 & 683 & 690 \\
192 & 1999 &  112 &  84 &  55 \\
196 & 1999 &  265 & 185 & 132
\end{tabular}
\end{center}
\vspace*{-0.7cm}
\end{table}

The search for Higgs bosons is performed in different analysis
channels with various signatures defined by the decay properties
of the Higgs and Z bosons. 
The production and decay rates are mainly calculated
in the framework of three theoretical models: the Standard
Model with one Higgs doublet; the general two-doublet Higgs model
where three neutral and two charged Higgs bosons are predicted;
and the MSSM, where in addition, Higgs boson masses and couplings 
are related. In the MSSM one neutral Higgs boson is predicted
to be light ($m_{\rm h}<130$ GeV), while the charged Higgs bosons
are heavier than the W boson.


\section{Detectors}
\begin{figure}[hp]
\caption{\label{fig:delphi} A typical LEP detector.}
\vspace*{-2mm}
\begin{center}
\includegraphics[width=\textwidth]
{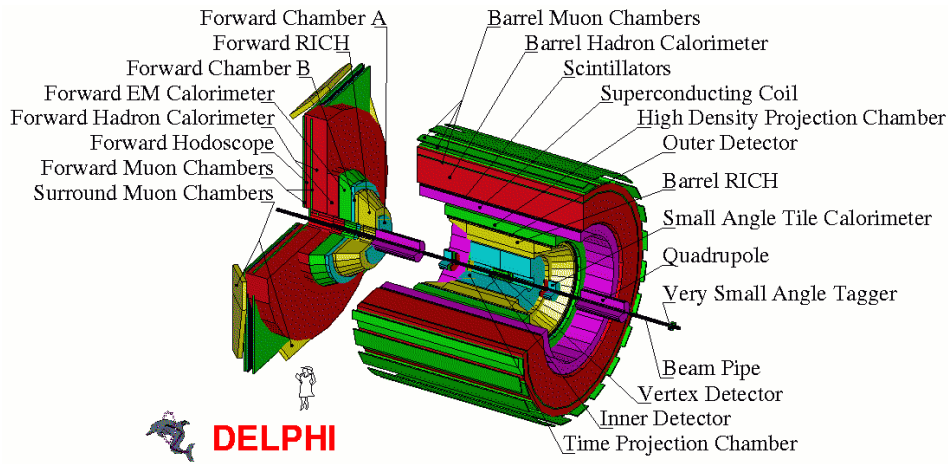}
\end{center}
\end{figure}

\begin{figure}[hp]
\caption{\label{fig:ident} Principle of particle identification.}
\vspace*{-2mm}
\begin{center}
\includegraphics[width=0.7\textwidth]{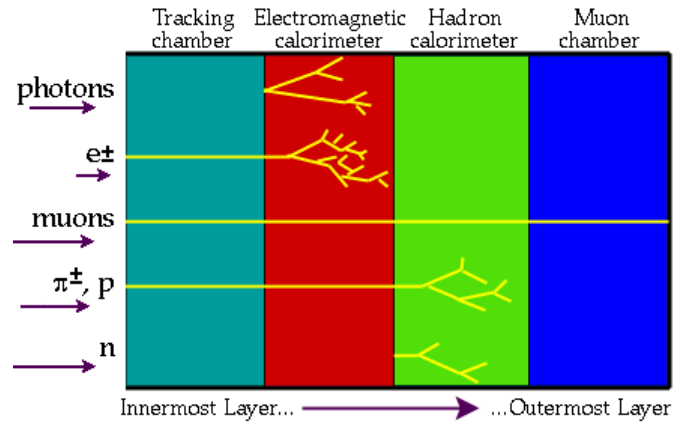}
\end{center}
\vspace*{-1cm}
\end{figure}

A typical LEP detector is shown in detail in Fig.~\ref{fig:delphi}.
The particles produced at the interaction point in the center
of the detector pass several subdetectors. 
All LEP detectors are equipped with micro-vertex detectors in order
to measure tracks with precision near the interaction point. Following
particles to larger radii, they pass a tracking system, 
electromagnetic calorimeters, hadron calorimeters and muon chambers.
Figure~\ref{fig:ident} illustrates the principle of the particle 
identification. Detailed particle identification is also possible 
with the DELPHI RICH detector using Cherenkov light to determine
particle masses.

\clearpage
\section{ZZ Background and Higgs Boson Candidates}
\begin{figure}[hp]
\caption{\label{fig:zz} Background production cross sections (left) and
          mass spectrum of ZZ events for data, 
          ZZ and background simulation (right).}
\vspace*{-35mm}
\begin{center}
\includegraphics[width=0.49\textwidth]{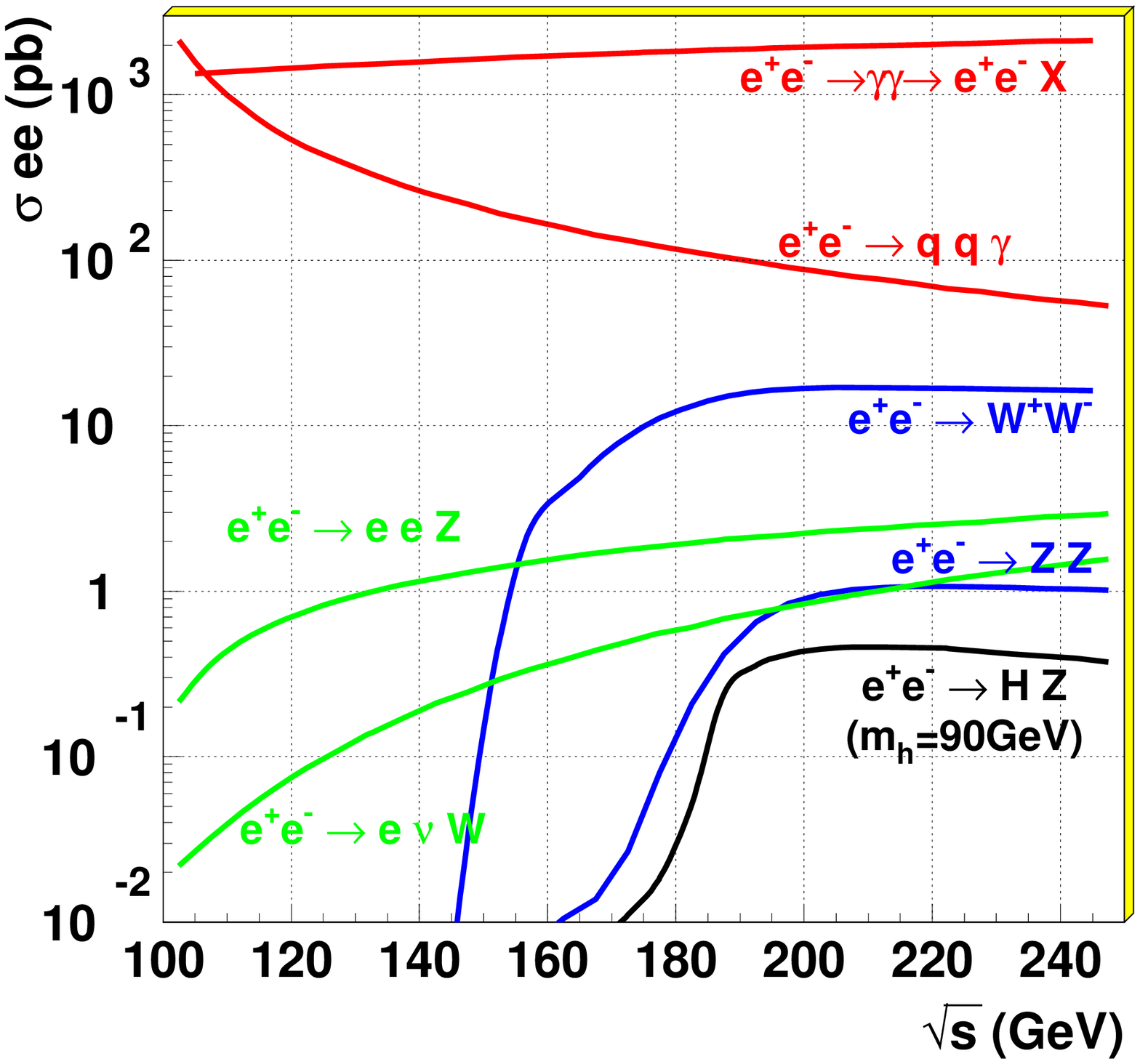}
\includegraphics[width=0.38\textwidth]{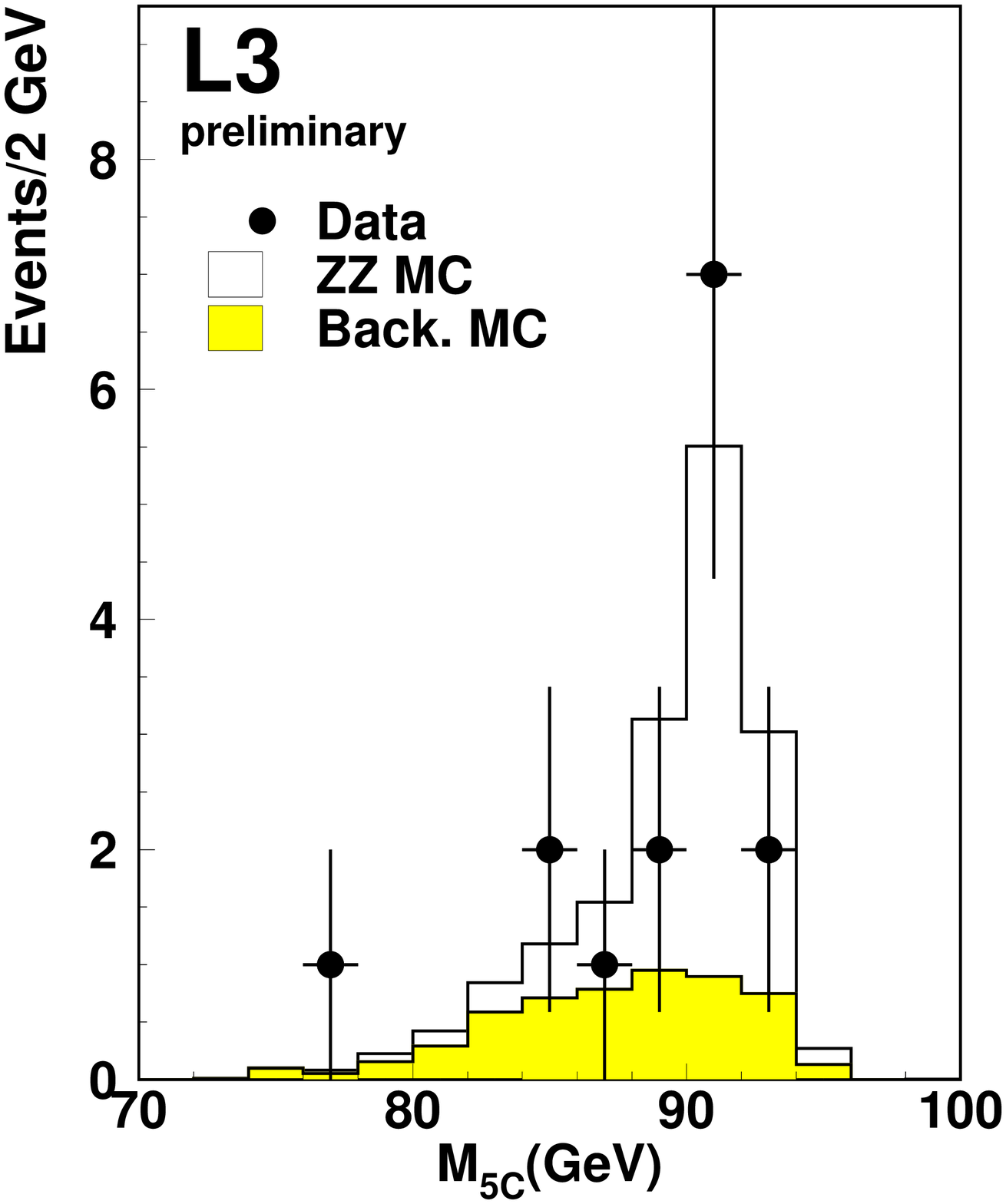}
\end{center}
\vspace*{-2.5cm}
\end{figure}

\begin{figure}
\vspace*{-0.3cm}
\begin{minipage}{0.30\textwidth}
\caption{\label{fig:btag} Expected background as a function of
         efficiency for a simulated hZ and hA signal.
The tagging of b-quarks reduces very strongly the WW process, where almost no 
b-quark decays are expected.
Also, qq and ZZ processes are largely suppressed.
While the Higgs bosons decay with more than about 85\% branching ratio 
into b-quark pairs, the $\rm Z\rightarrow bb$ decay rate is only 15\%.}
\end{minipage}
\hfill
\begin{minipage}{0.62\textwidth}
\vspace*{-0.5cm}
\begin{center}
\includegraphics[width=1.0\textwidth,%
bbllx=29pt,bblly=164pt,bburx=530pt,bbury=675pt,clip=]
{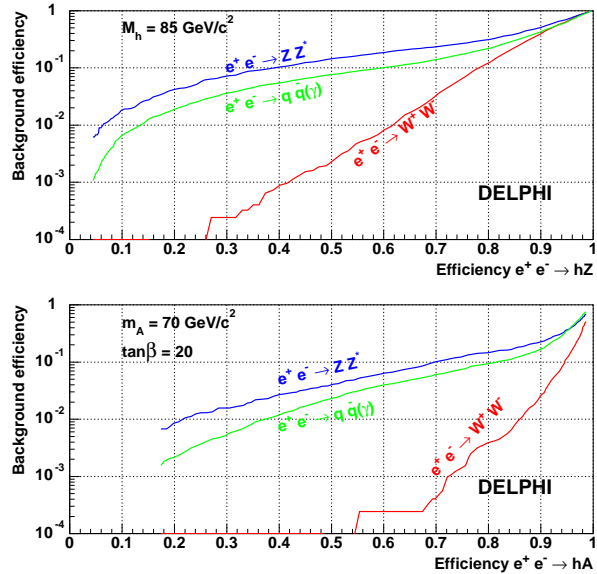}
\end{center}
\end{minipage}
\vspace*{-0.2cm}
\end{figure}

The production cross sections for the background 
reactions are shown in Fig.~\ref{fig:zz}.
In 1998, the threshold of the ZZ production was passed and 
the ZZ production and decay became a dominant 
background for many searches.
Thus, a good understanding of this background is
very important to observe a signal.
The good agreement of data and simulation is shown in
Fig.~\ref{fig:zz} for the L3 
$\rm e^+e^-\rightarrow ZZ\rightarrow qq\ell\ell$ analysis~\cite{l3zz}.
An important method to reduce the ZZ background is based
on the identification of b-quarks. 
As an example, the b-tagging 
performance from DELPHI is shown in Fig.~\ref{fig:btag}.

\begin{figure}[hp]
\caption{\label{fig:bbbb} DELPHI four-jet Higgs boson candidate event.}
\begin{center}
\includegraphics[width=1.0\textwidth]{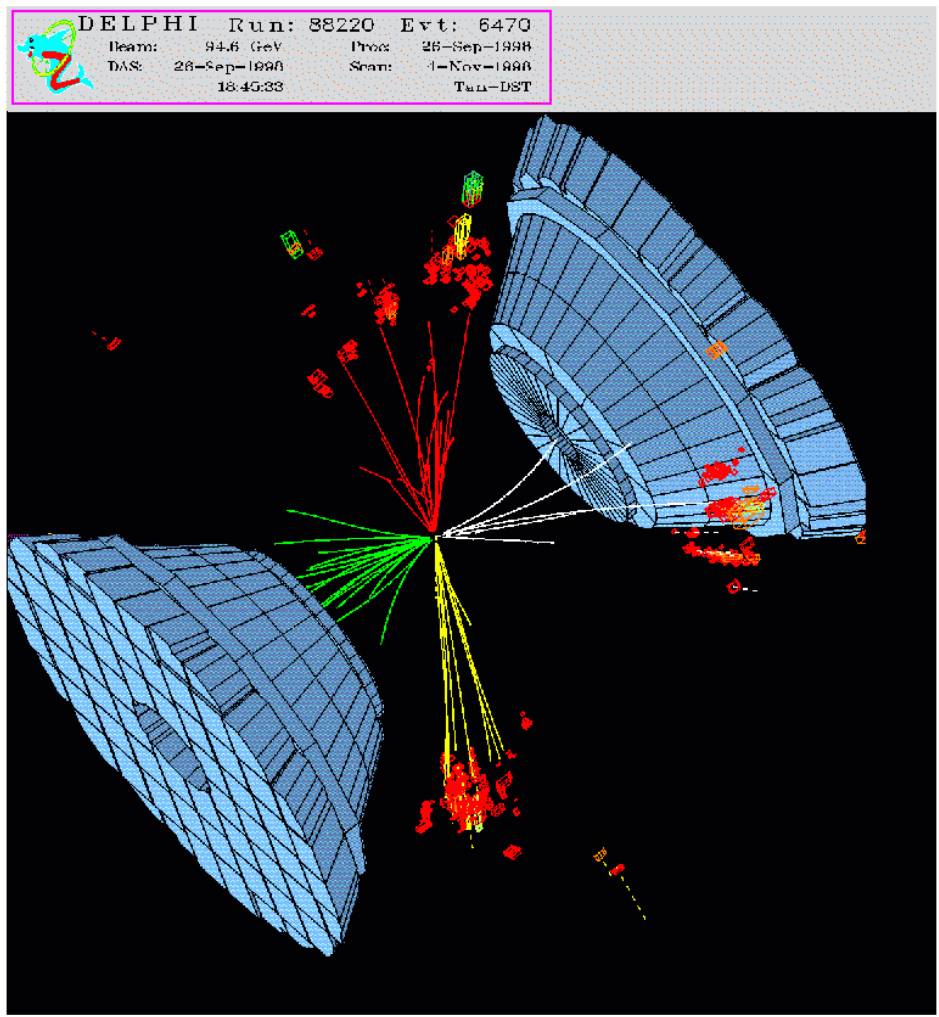}
\end{center}
\vspace*{-1.5cm}
\end{figure}

\begin{figure}[hp]
\caption{\label{fig:qqee} ALEPH jet--jet $\rm e^+e^-$ Higgs boson 
candidate event.
The upper right-hand figure shows nicely the separation of the $\rm e^+e^-$ 
pair from the two hadronic jets,
while the lower figures show clearly the displaced vertex from
a B-meson decay.}
\begin{center}
\epsfig{figure=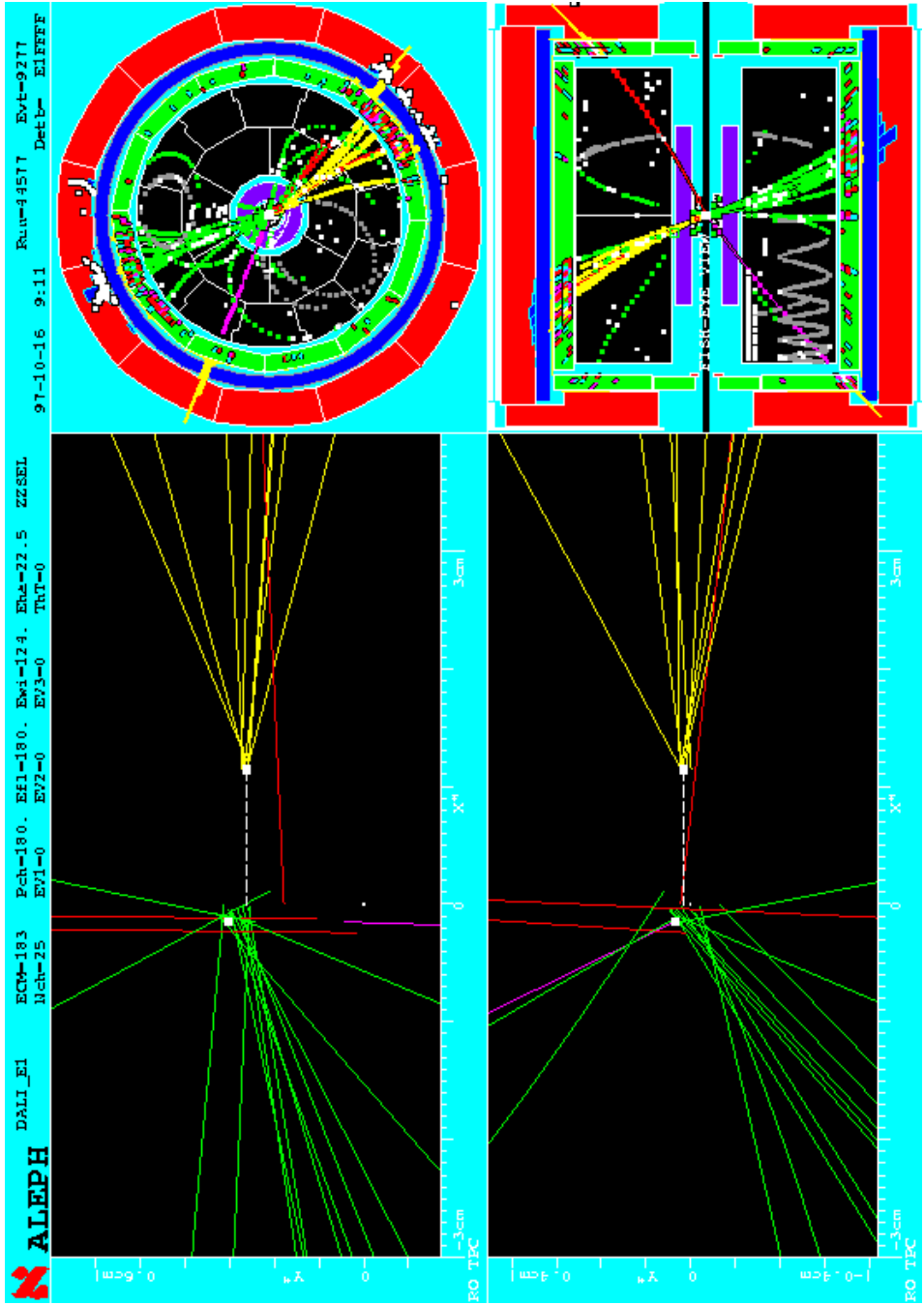,width=\textwidth}
\end{center}
\vspace*{-1.5cm}
\end{figure}

\begin{figure}[hp]
\caption{\label{fig:qqmm} DELPHI jet--jet $\mu^+\mu^-$ Higgs boson 
candidate event.}
\begin{center}
\includegraphics[width=1.0\textwidth]{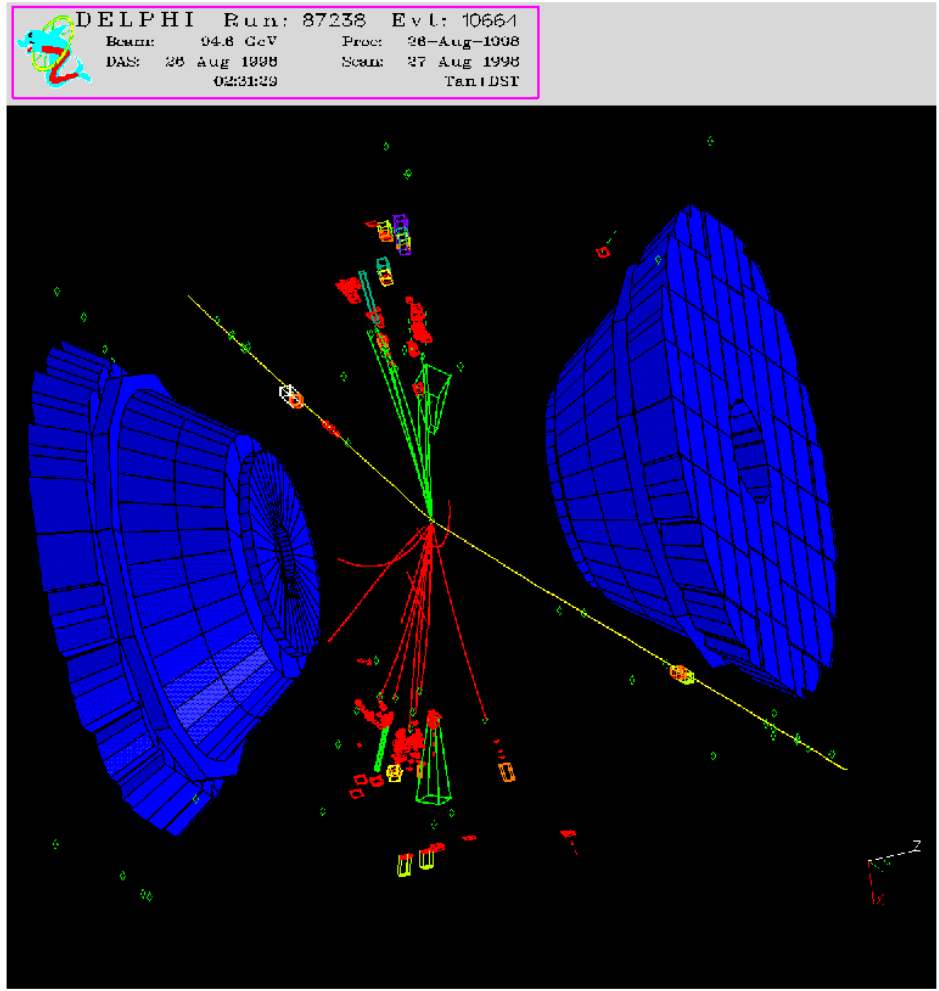}
\end{center}
\vspace*{-0.5cm}
\end{figure}

\clearpage
Figures~\ref{fig:bbbb},~\ref{fig:qqee} and~\ref{fig:qqmm} 
show examples 
of selected SM Higgs candidates.
The reconstructed invariant masses are 92.9 and 91.1~GeV
in the four-jet event (Fig.~\ref{fig:bbbb}).
A typical jet--jet $\rm e^+e^-$ event signature is shown
in Fig.~\ref{fig:qqee} where the invariant masses are 85 and 96 GeV.
One jet is b-tagged by a three dimensional reconstruction of a 
secondary vertex.
In the jet--jet $\mu^+\mu^-$ event (Fig.~\ref{fig:qqmm})
the invariant mass of the muon pair is 92.3 GeV.
The hadron invariant mass is calculated to be 93.4 GeV
from the recoil mass of the muon pair for a
better resolution.
All reconstructed masses are compatible with ZZ production,
taking into account the errors on the measurements.

\clearpage


\clearpage

\clearpage
\section{Standard Model Higgs Boson}

\begin{figure}[htbp]
\vspace*{-2mm}
\caption{\label{fig:msmspectra} 
Reconstructed mass distribution for combined data between 192 and
196 GeV (left) and including 189 GeV data (right).}
\begin{center}
\vspace*{-0.3cm}
\mbox{\epsfig{file=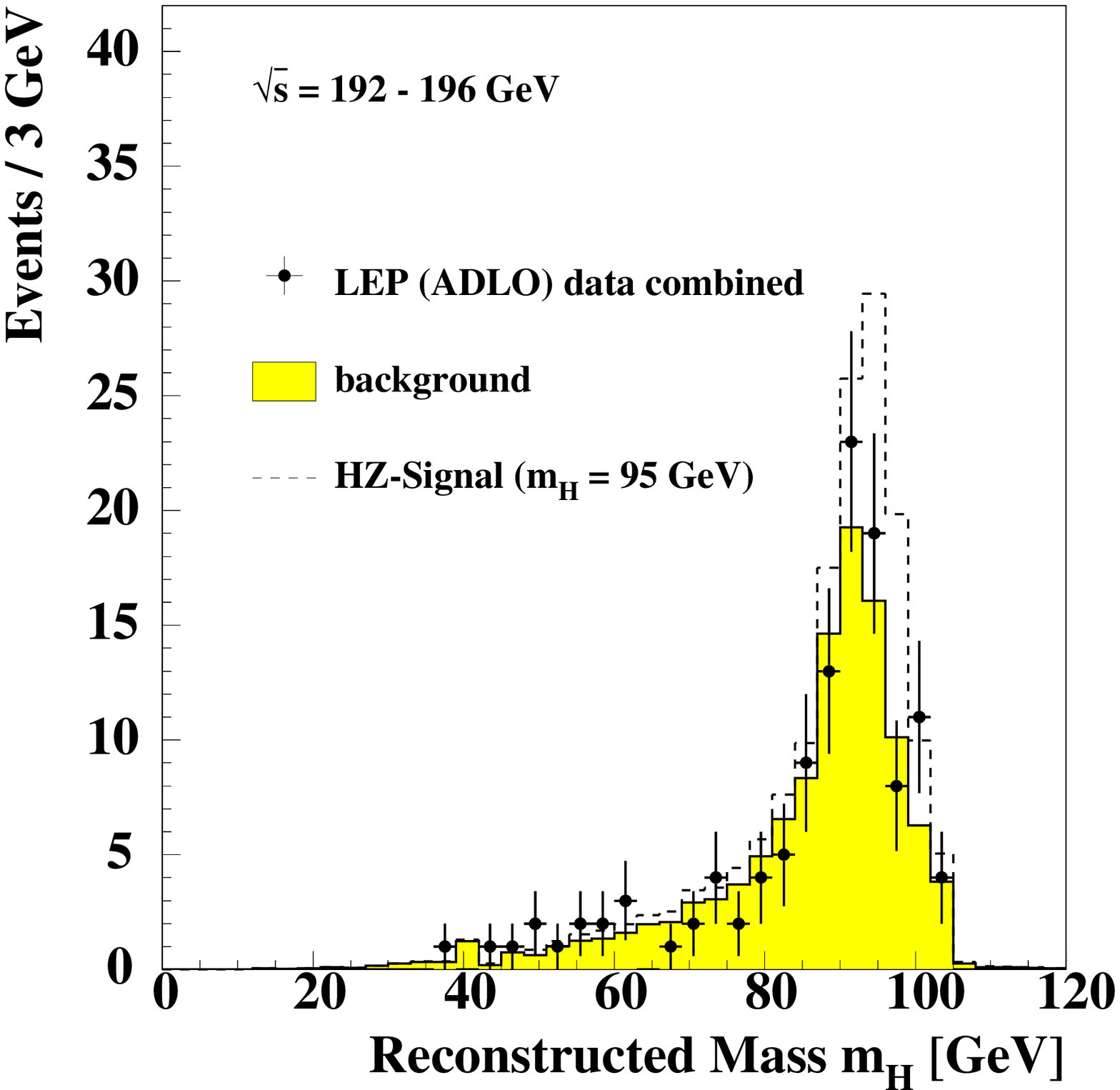,width=0.49\textwidth}}
\mbox{\epsfig{file=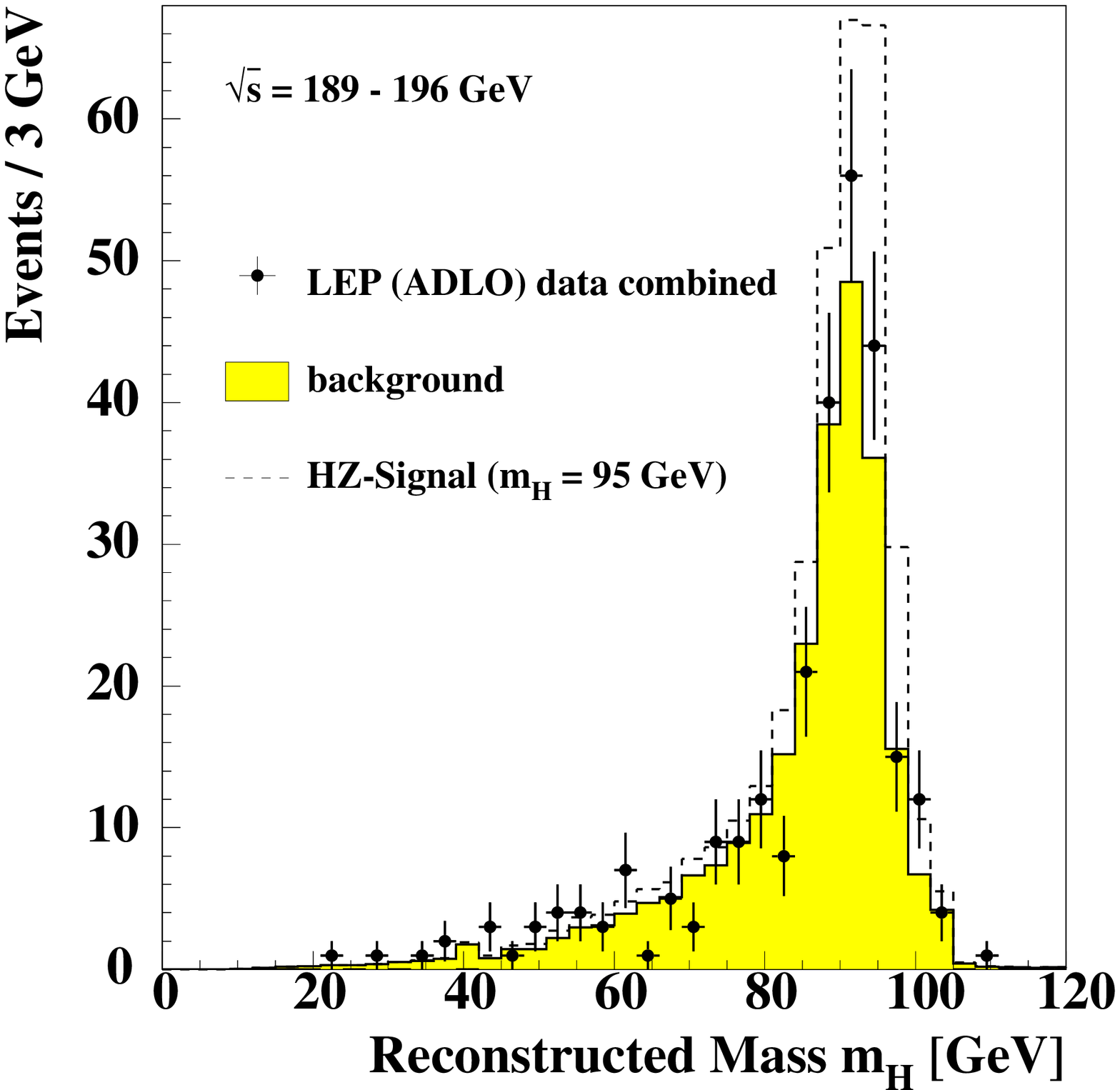,width=0.49\textwidth}}
\end{center}
\vspace*{-1cm}
\end{figure}

\begin{figure}[htbp]
\caption{\label{fig:msmclb} $1-CL_{\rm b}$ distribution for 1998 data
         (left) where a probability of 0.01 is observed for a
         96 GeV Higgs boson. The data up to 196 GeV (right)
         reject the possibility of a 96 GeV Higgs boson.
         The grey areas show 1 and $2\sigma$ regions.}
\begin{center}
\vspace*{-0.3cm}
\mbox{\epsfig{file=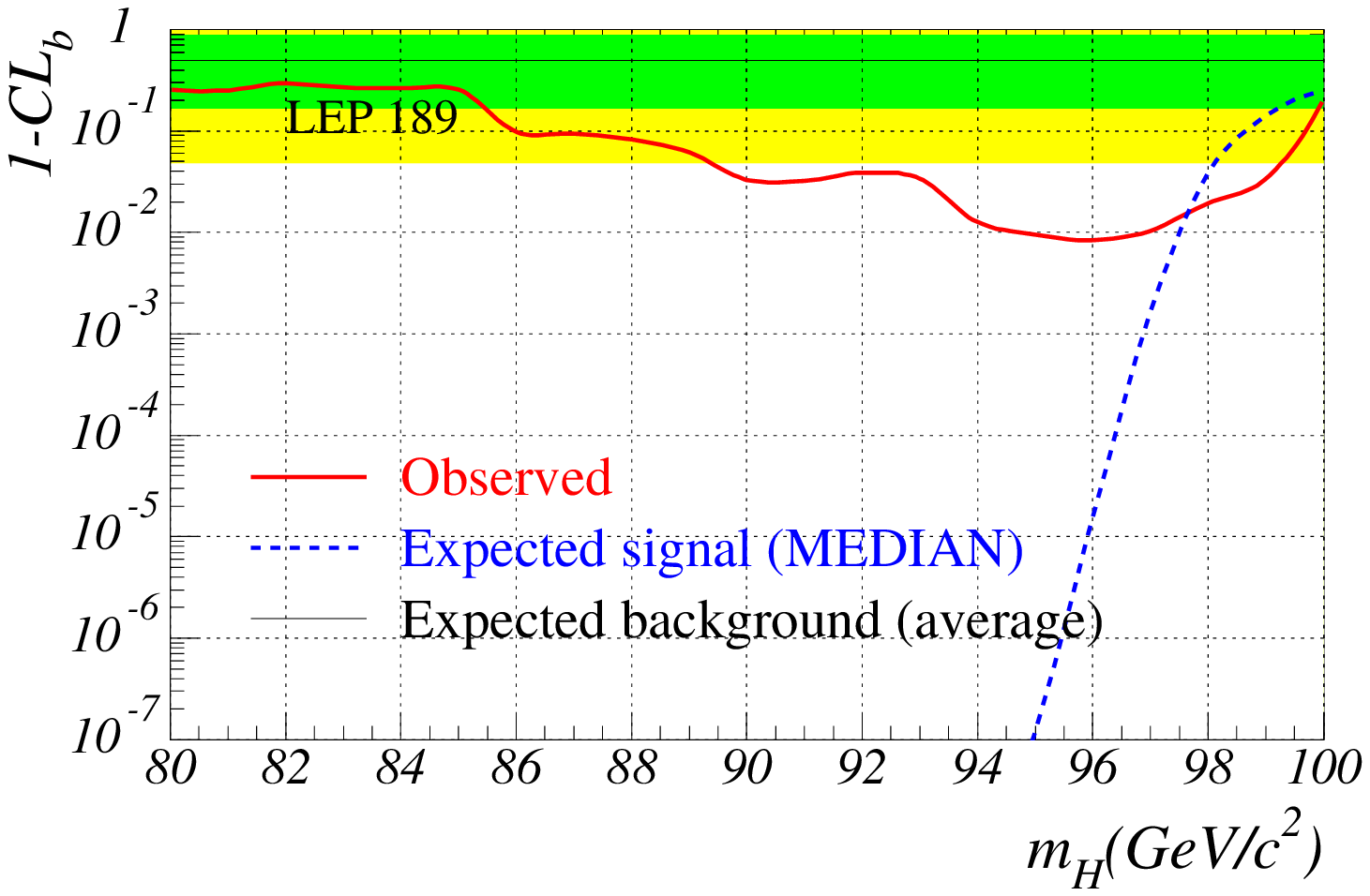,width=0.49\textwidth}}
\mbox{\epsfig{file=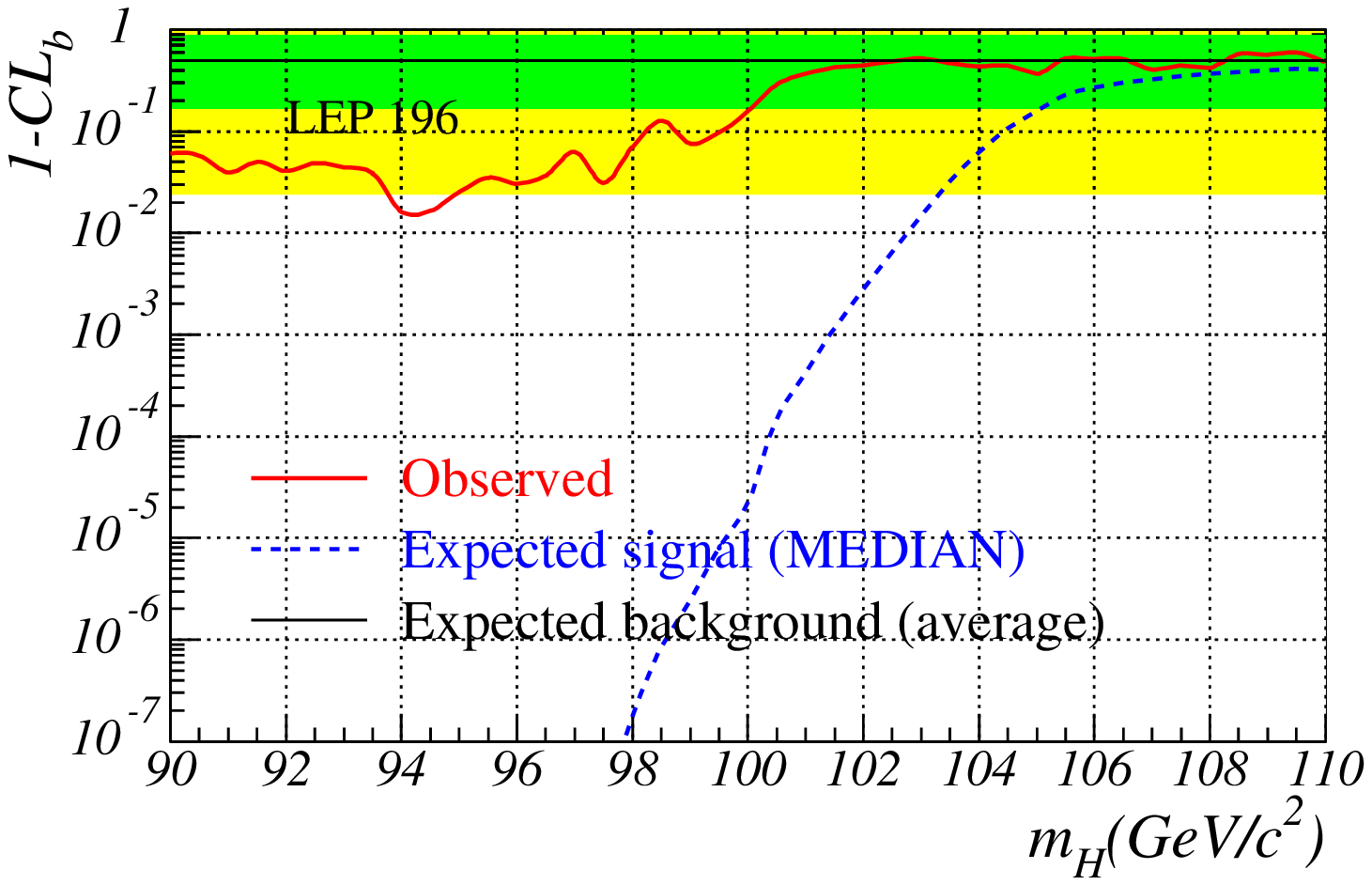,width=0.49\textwidth}}
\end{center}
\vspace*{-7mm}
\end{figure}

\begin{figure}[tp]
\vspace*{-3mm}
\caption{\label{fig:msmlimit}
Observed and expected SM Higgs boson mass limits, where 95\% CL
corresponds to $CL_{\rm s} = 0.05$ (left)
and $\Delta\chi^2$ values for a given Higgs boson mass (right).
The grey areas show 1 and $2\sigma$ regions.}
\begin{center}
\vspace*{-0.3cm}
\mbox{\epsfig{file=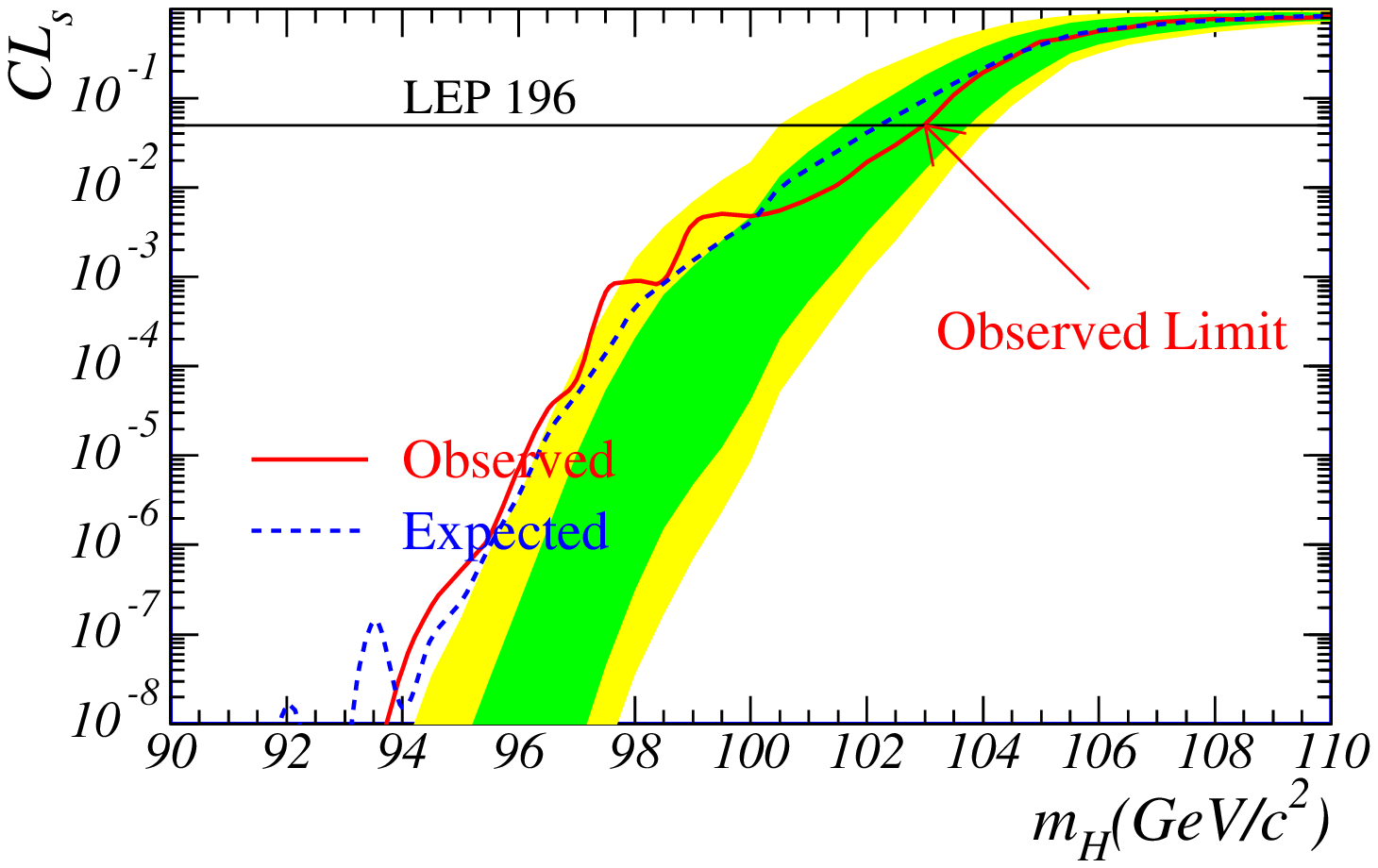,width=0.49\textwidth}}
\mbox{\epsfig{file=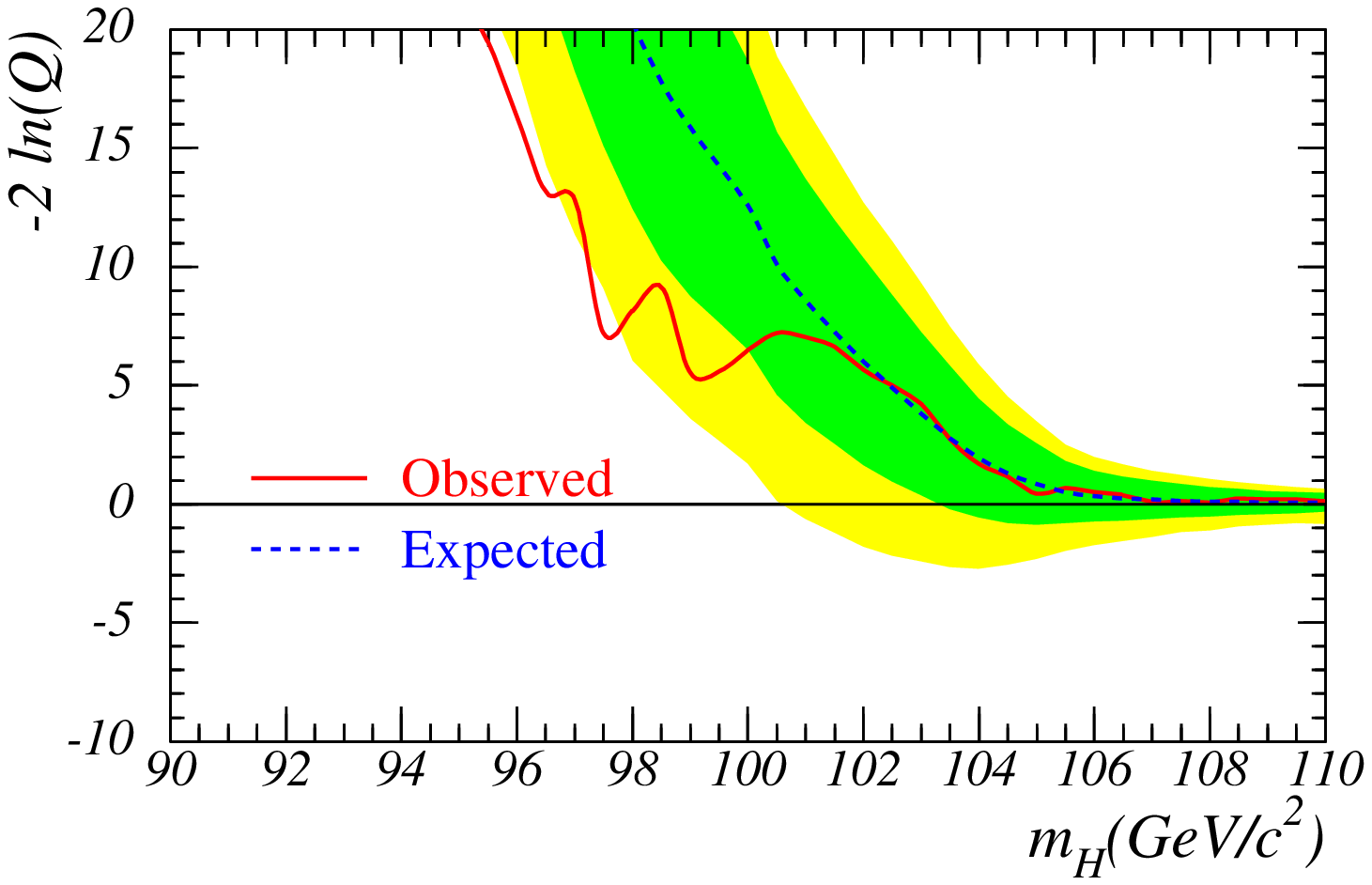,width=0.49\textwidth}}
\end{center}
\vspace*{-0.7cm}
\end{figure}

The search for the Standard Model Higgs boson is performed in the 
following decay channels: 
$\rm HZ\rightarrow~bbqq,~bbee,~bb\mu\mu,~bb\tau\tau~and~bb\nu\nu$.
Figure~\ref{fig:msmspectra} shows the number of data and background 
events as a function of the reconstructed Higgs boson mass. 
No indication of a signal is observed.
A statistical analysis based on a likelihood function takes
into account all search channels and 
gives the probability $1-CL_{\rm b}$ that the data are compatible with the 
simulated background. Figure~\ref{fig:msmclb} shows this
probability for 1998 data and including 1999 data up to 196 GeV
center-of-mass energy.
The limits on the SM Higgs boson mass derived from the probability
$CL_{\rm s}$ that the expected signal and background is compatible 
with the data, and the corresponding $\Delta\chi^2$ distribution 
are presented in Fig.~\ref{fig:msmlimit}.

\clearpage

\begin{figure}[htbp]
\caption{\label{fig:xi} 
Combined limit on the HZZ suppression factor $\xi^2$ 
for LEP1 and LEP2 data up to 196 GeV center-of-mass energy (left)
and DELPHI 189 GeV invisible Higgs limits in the plane of
mass and mixing angle for H and S bosons (right).}
\begin{center}
\vspace*{-0.4cm}
\includegraphics[width=0.49\textwidth]{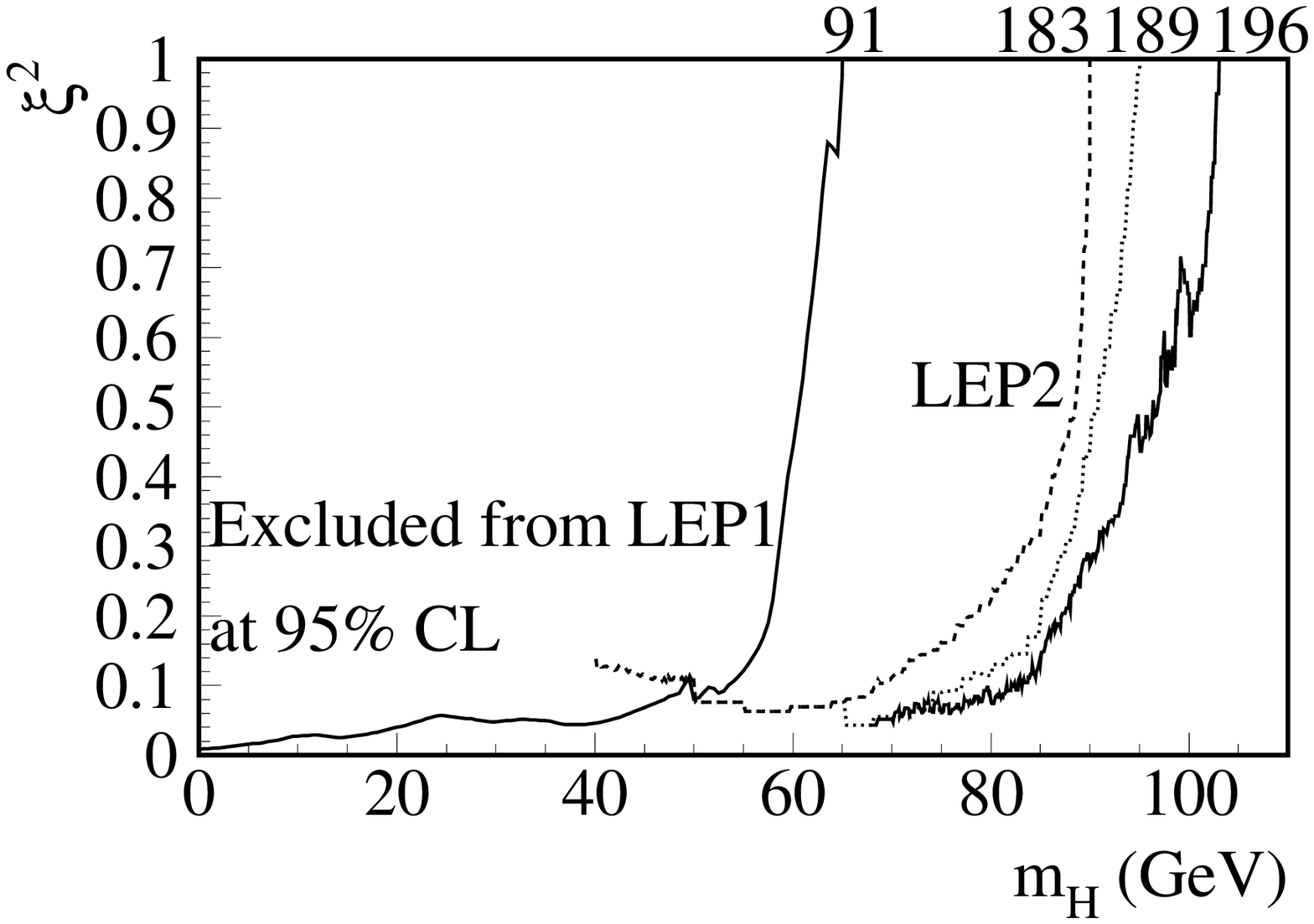}
\includegraphics[width=0.45\textwidth]{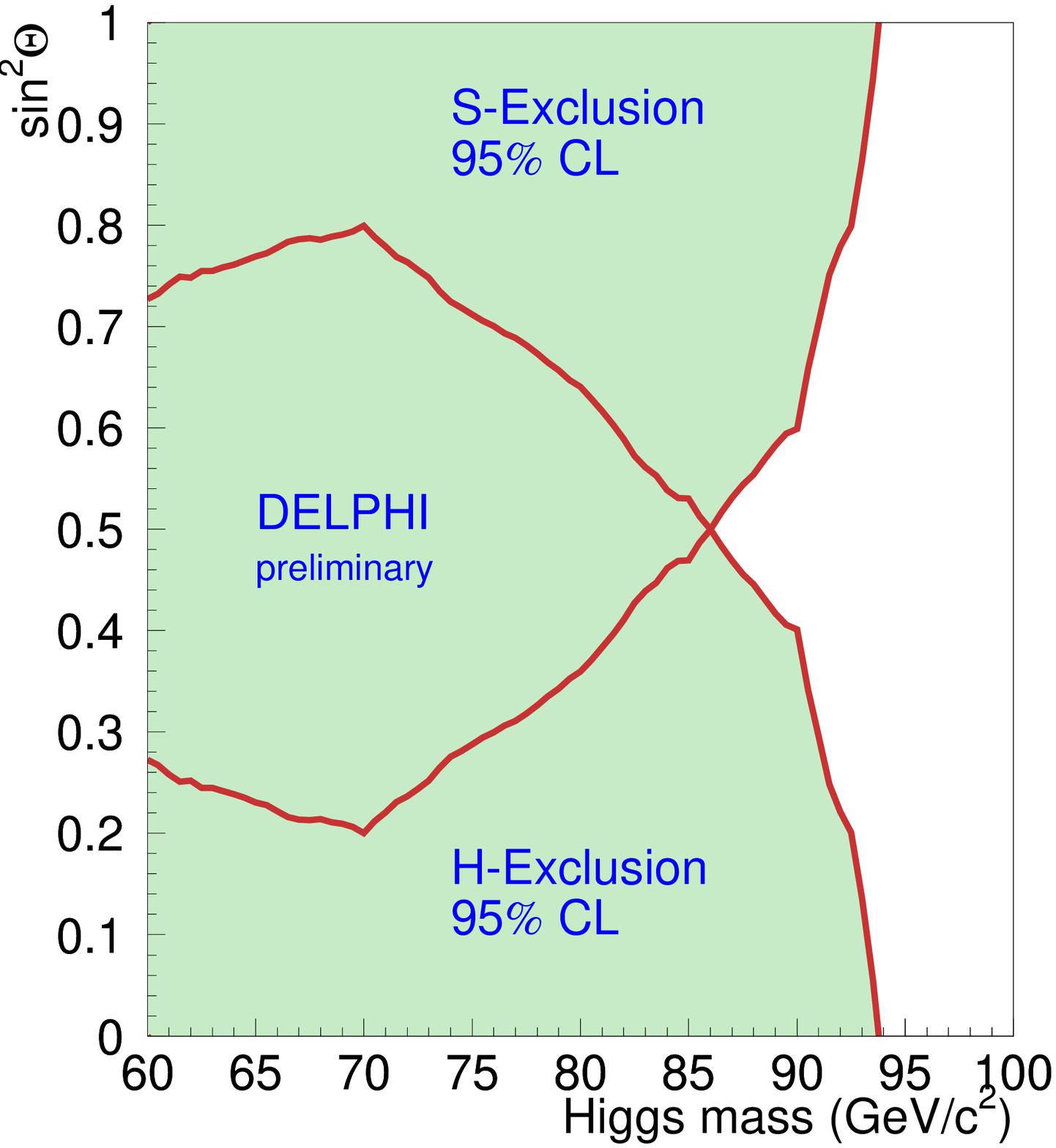}
\end{center}
\vspace*{-1cm}
\end{figure}

The HZZ coupling is predicted to be smaller in many extensions of 
the SM.
Figure~\ref{fig:xi} shows stringent limits on $\xi^2$
from combined LEP data under the assumption that the Higgs boson 
decay modes do not change and the production rate is given by
$\rm \sigma(HZ) = \sigma(H_{SM} Z) \cdot \xi^2$.


An interesting extension of the SM is the addition of 
an extra Higgs singlet field. In this case two neutral
massive Higgs bosons H and S are predicted, as is a
massless, stable and non-interacting Majoron, J. The 
decays $\rm H\rightarrow JJ$ and $\rm S\rightarrow JJ$
lead to missing energy signatures. This process could 
be observed in the ZH and ZS production.
The associated Z decay into a quark pair gives a signature
of two jets and missing energy as in the H$\nu\nu$ search channel.
A limit from DELPHI on the mixing angle of H and S as a 
function of their masses is shown in Fig.~\ref{fig:xi}.


\clearpage
\section{MSSM Higgs Bosons}

\begin{figure}[hp]
\caption{\label{fig:mssmsprectrum} 
Reconstructed mass distribution for combined data between 192 and
196 GeV (left) and including 189 GeV data (right).}
\begin{center}
\mbox{\epsfig{file=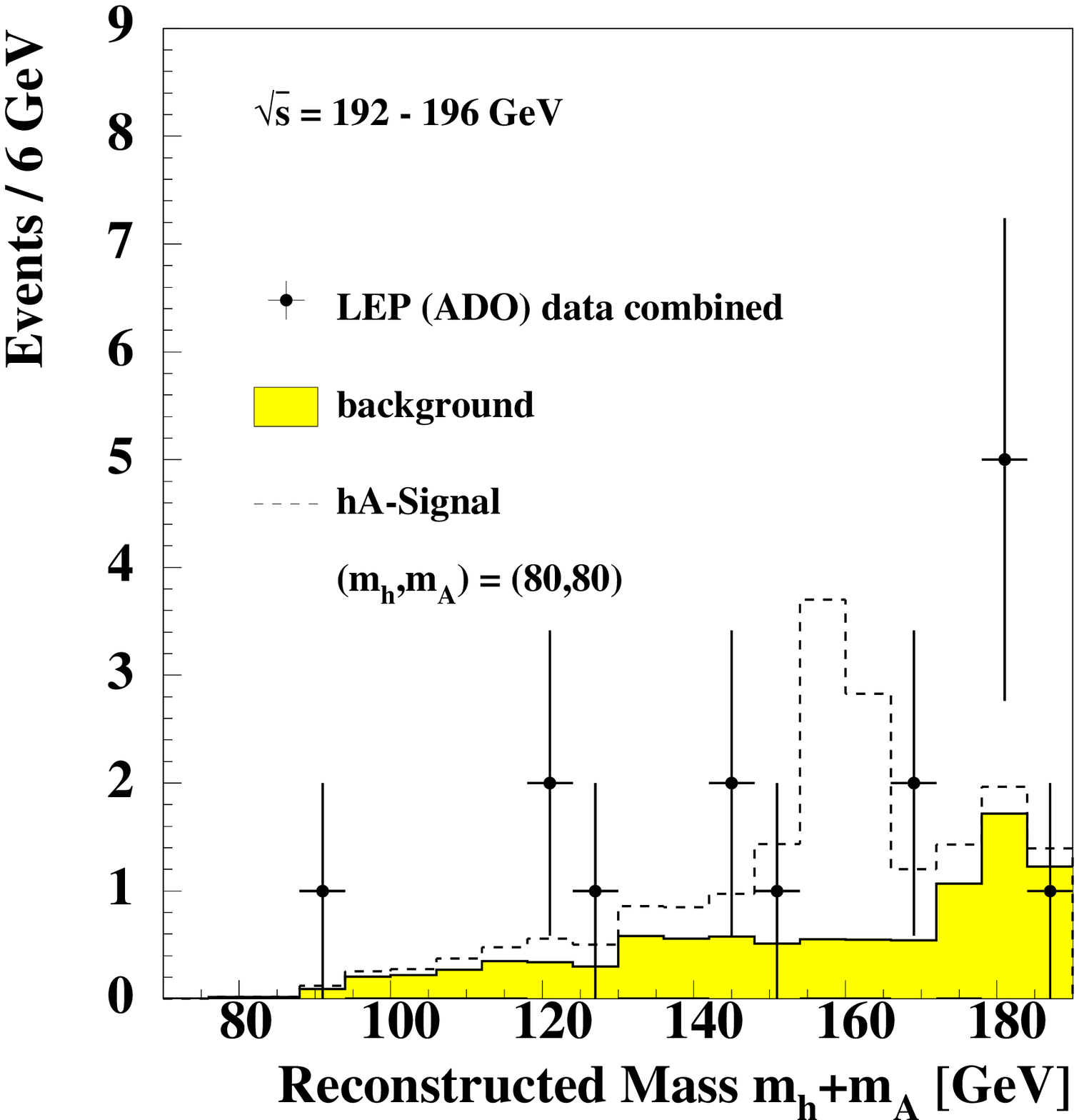,width=0.49\textwidth}}
\mbox{\epsfig{file=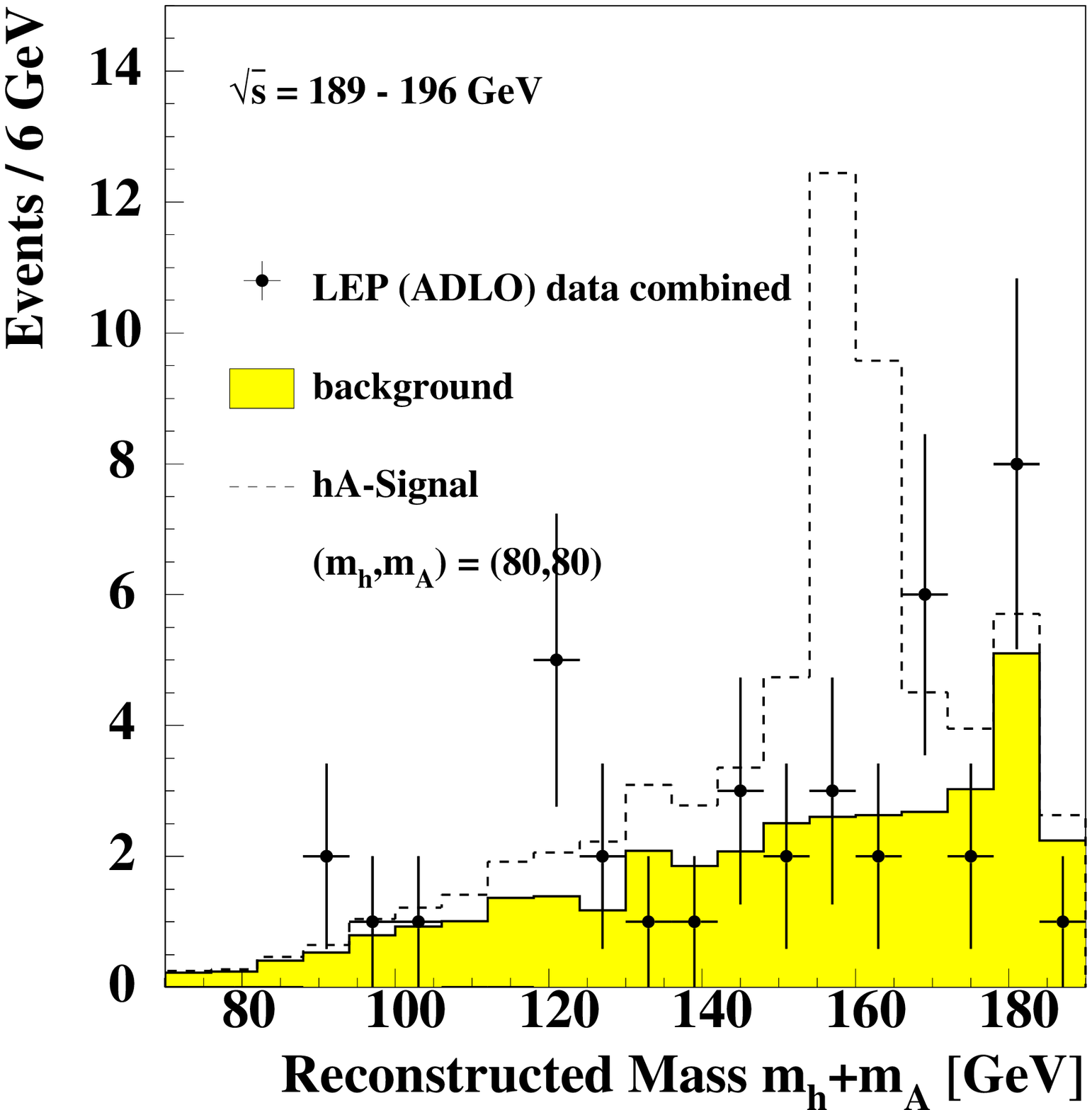,width=0.49\textwidth}}
\end{center}
\end{figure}

The search for the scalar MSSM Higgs bosons is performed
in the same search channels as for the SM Higgs boson.
In addition, the pair-production process 
$\rm e^+e^-\rightarrow hA\rightarrow bbbb~and~\tau\tau bb$ is investigated,
where h is the lighter scalar and A the pseudoscalar Higgs boson
of the MSSM.
Figure~\ref{fig:mssmsprectrum} shows the
sum of the reconstructed Higgs boson masses.
For simulated Higgs boson masses of 80 GeV,
well below the kinematic production threshold, 
the inclusion of the data-taken in 1998 at 189 GeV 
increases significantly the discovery sensitivity.

\begin{figure}[hp]
\caption{\label{fig:matb}
MSSM benchmark exclusion in the ($m_{\rm A},\tan\beta$) plane.}
\begin{center}
\vspace*{-0.5cm}
\mbox{\epsfig{file=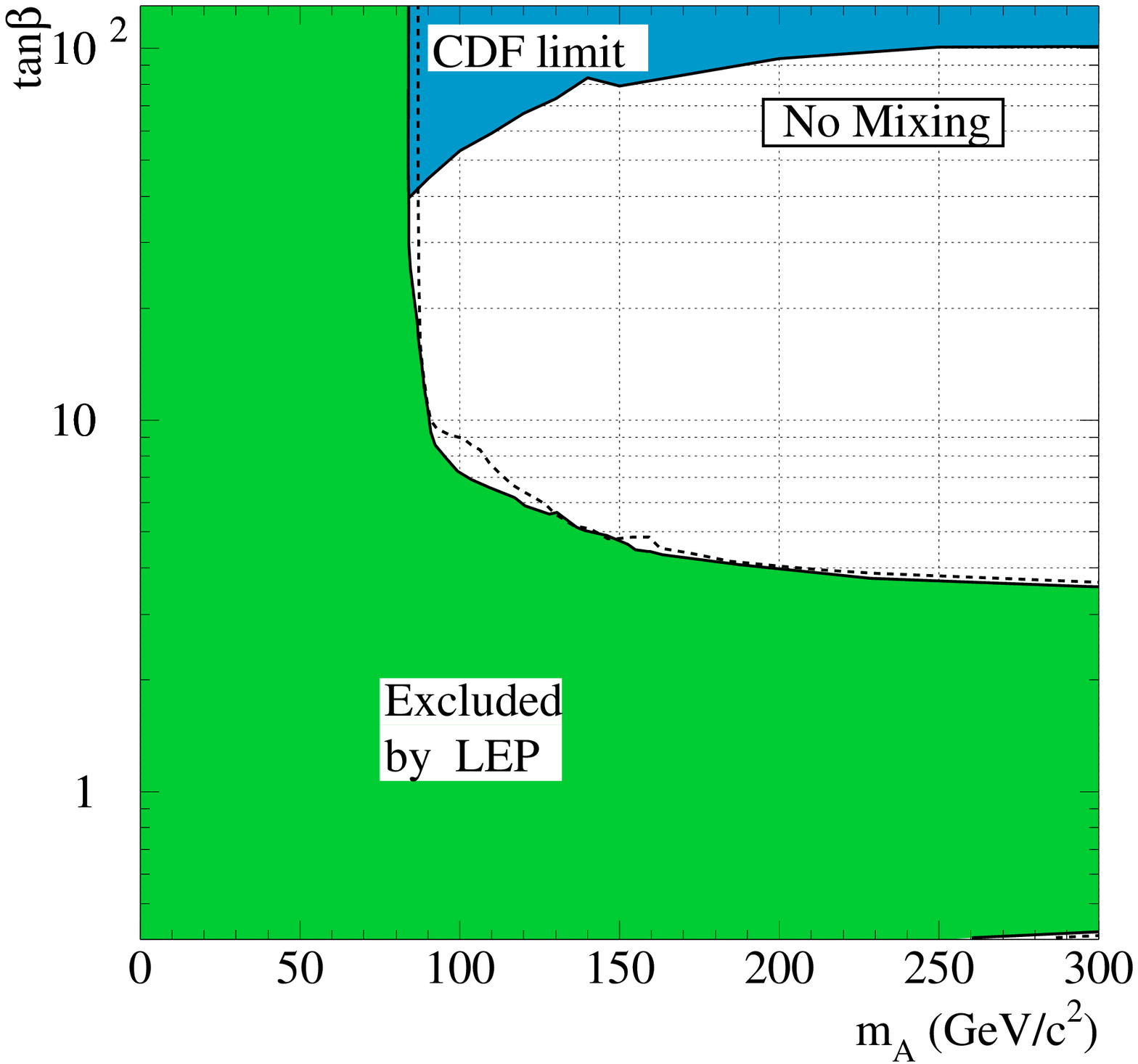,width=0.49\textwidth}}
\mbox{\epsfig{file=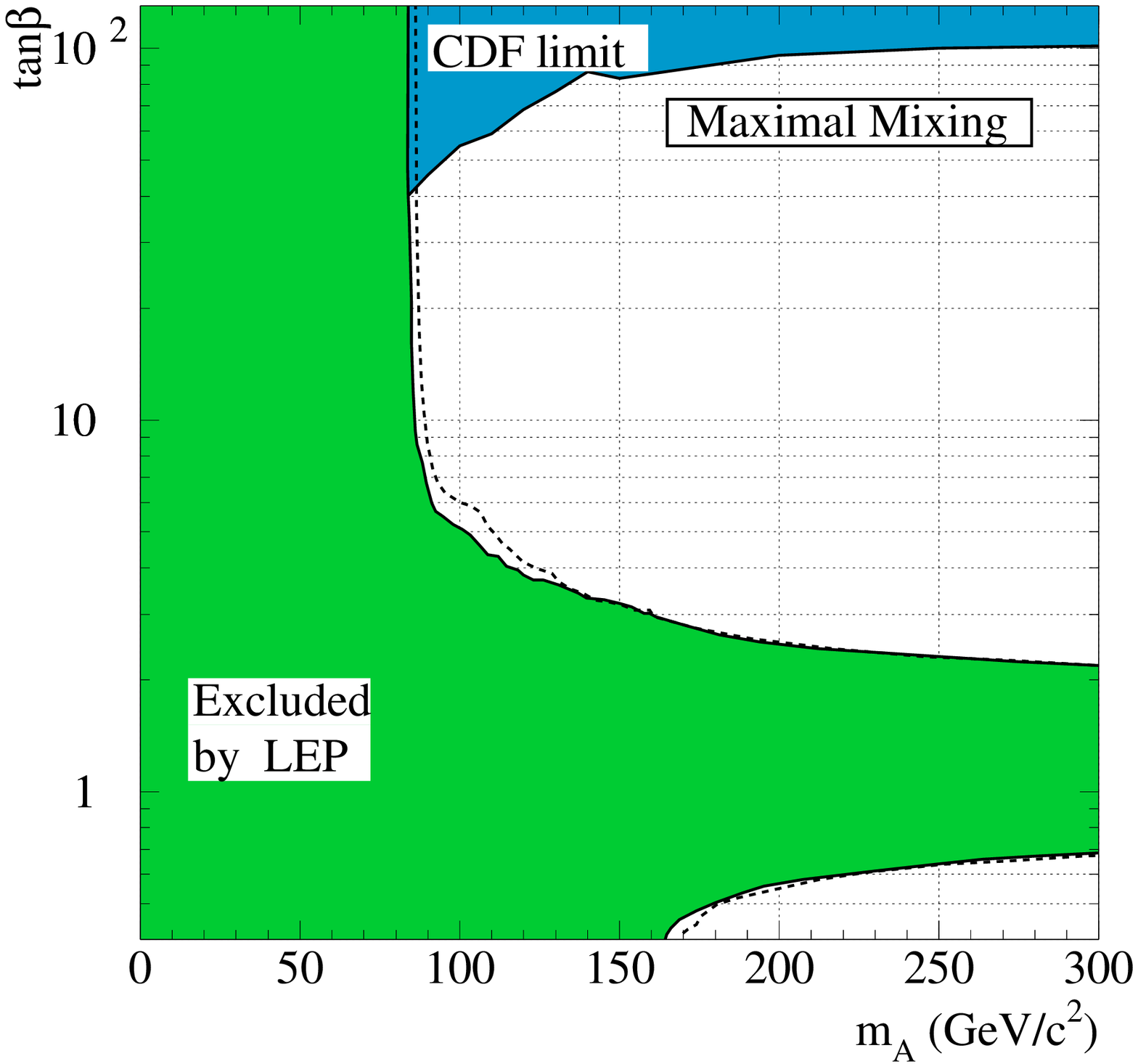,width=0.49\textwidth}}
\end{center}
\vspace*{-0.9cm}
\end{figure}

\begin{figure}[hp]
\caption{\label{fig:mhtb}
MSSM benchmark exclusion in the ($m_{\rm h},\tan\beta$) plane.}
\begin{center}
\vspace*{-0.4cm}
\mbox{\epsfig{file=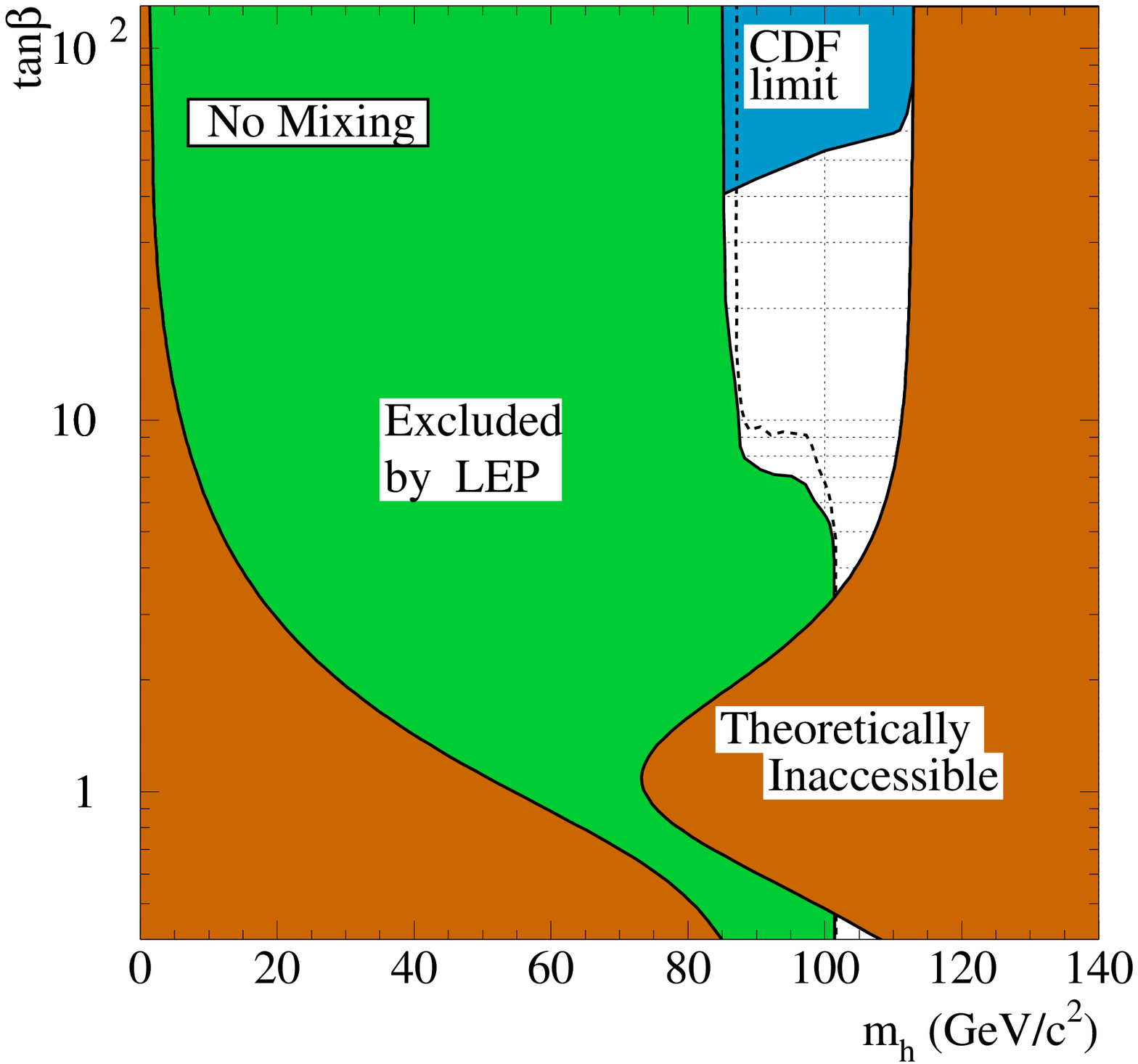,width=0.49\textwidth}}
\mbox{\epsfig{file=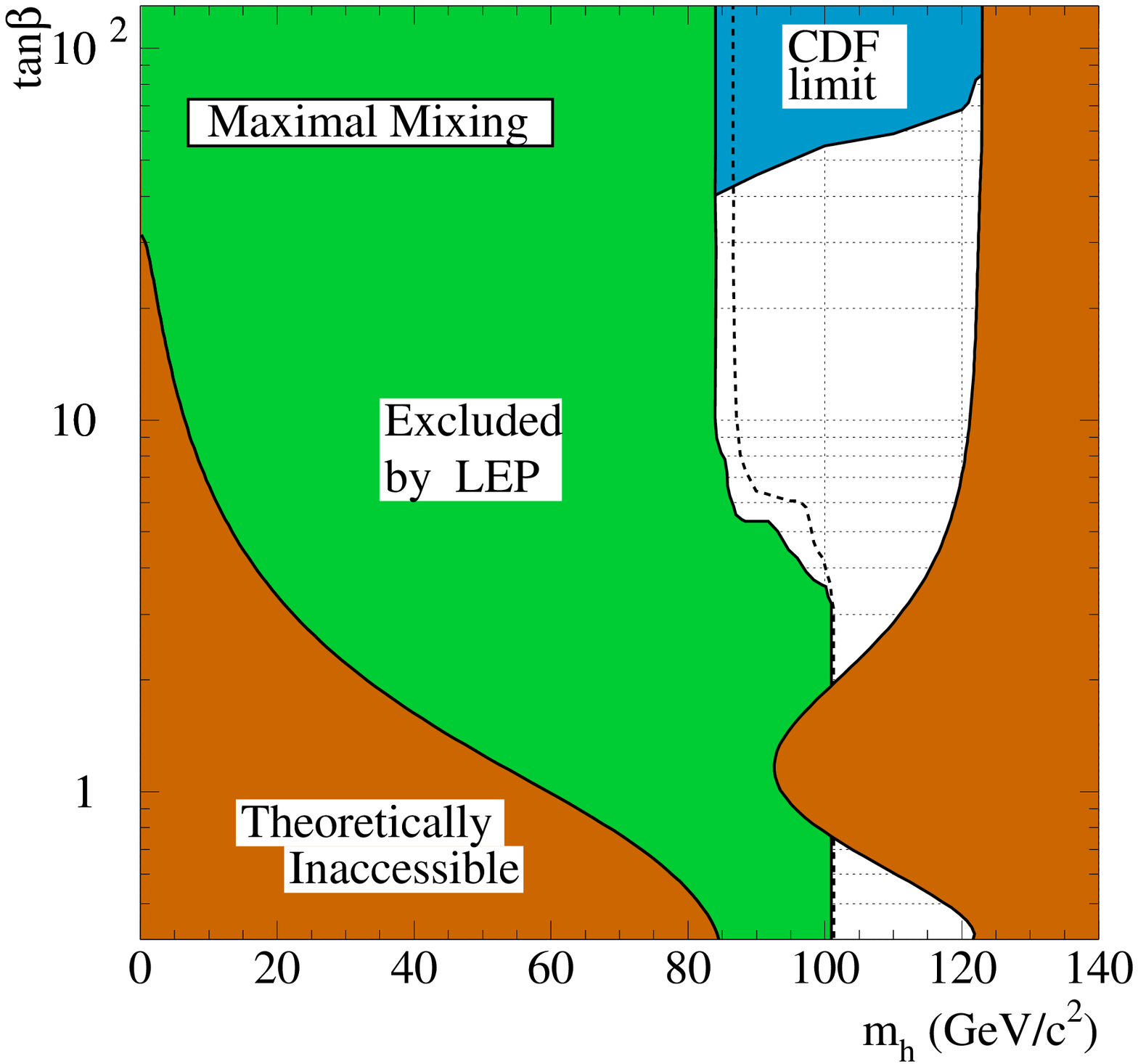,width=0.49\textwidth}}
\end{center}
\vspace*{-0.9cm}
\end{figure}

\begin{figure}[htbp]
\caption{\label{fig:mssmdiscovery} MSSM benchmark
                      $1-CL_{\rm b}$ (left) and $CL_{\rm s}$ (right)
                      in the ($m_{\rm h},m_{\rm A}$) plane.}
\begin{center}
\vspace*{-0.3cm}
\mbox{\epsfig{file=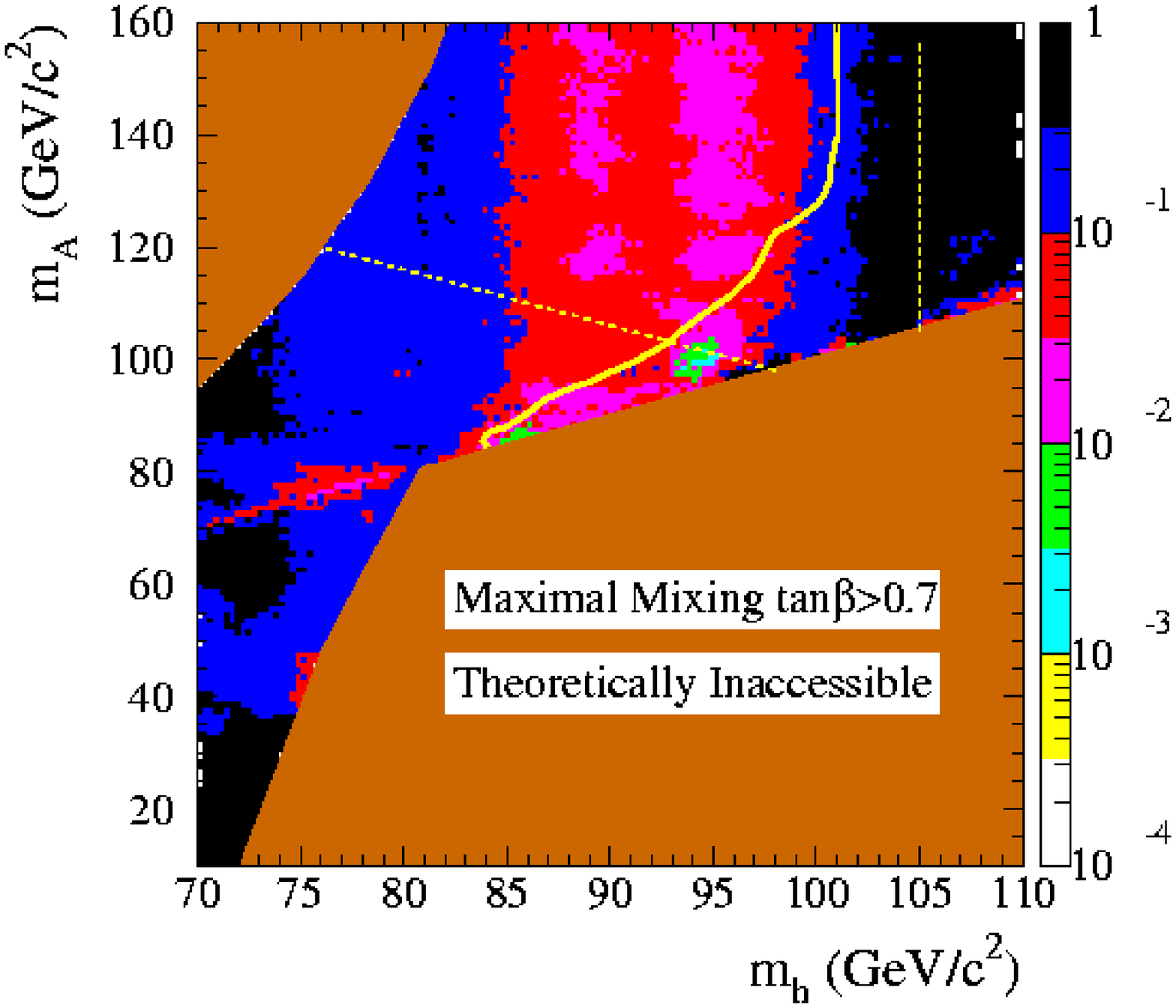,width=0.49\textwidth}}
\mbox{\epsfig{file=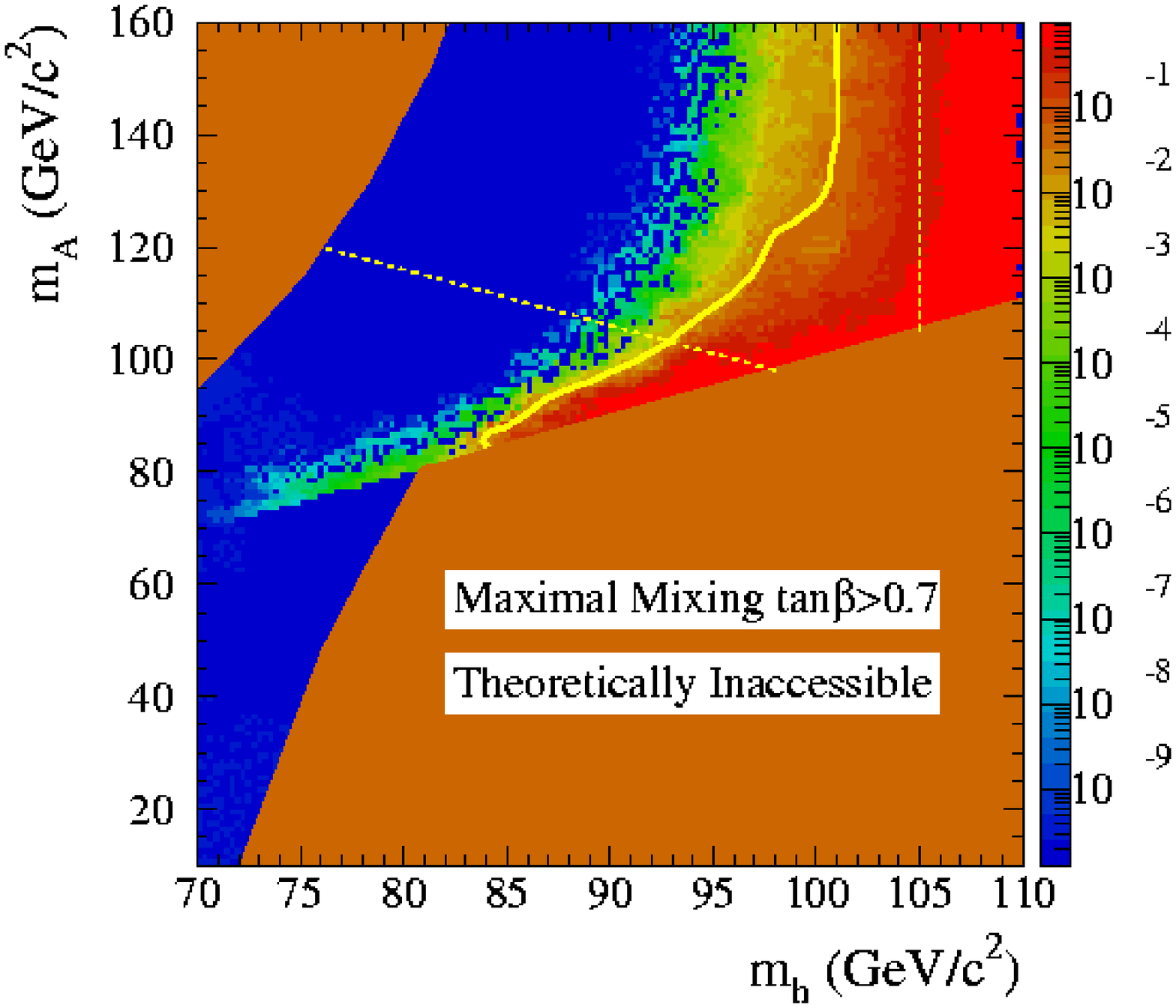,width=0.49\textwidth}}
\end{center}
\vspace*{-1.3cm}
\end{figure}

Interpretations are given first for the so-called
benchmark parameters, defined by a large Supersymmetric particle
scale of 1 TeV.
Figure~\ref{fig:matb} shows the excluded regions in
the $(m_{\rm A},\tan\beta)$ plane for no and maximal
mixing in the scalar top sector, and
Fig.~\ref{fig:mhtb} in the $(m_{\rm h},\tan\beta)$ plane.
The figures also show limits from the CDF experiment~\cite{cdf},
based on the reaction $\rm bb\rightarrow bbh$, where the
coupling is enhanced for large $\tan\beta$.
Based on the benchmark parameters, details of the discovery
($1-CL_{\rm b}$) and the exclusion ($CL_{\rm s}$) probabilities are
shown in Fig.~\ref{fig:mssmdiscovery}. The white line gives the 
95\% CL exclusion limit from $CL_{\rm s}$ in both plots
and the dashed lines the kinematic
production limits for hA and hZ processes.
No indication of a discovery is observed, where a $5\sigma$ 
effect corresponds to $1-CL_{\rm b} \approx 10^{-7}$ and
the 95\% CL corresponds to $CL_{\rm s} = 0.05$.

\clearpage

\begin{figure}[hp]
\vspace*{-1mm}
\caption{\label{fig:delphi189scan} DELPHI general scan leading to 
reduced mass limits compared to the benchmark results (left),
and comparison of 1- and 2-loop FDC as a function of the mixing $A$
in the scalar top sector (right).}
\begin{center}
\vspace*{-0.3cm}
\includegraphics[width=0.49\textwidth]{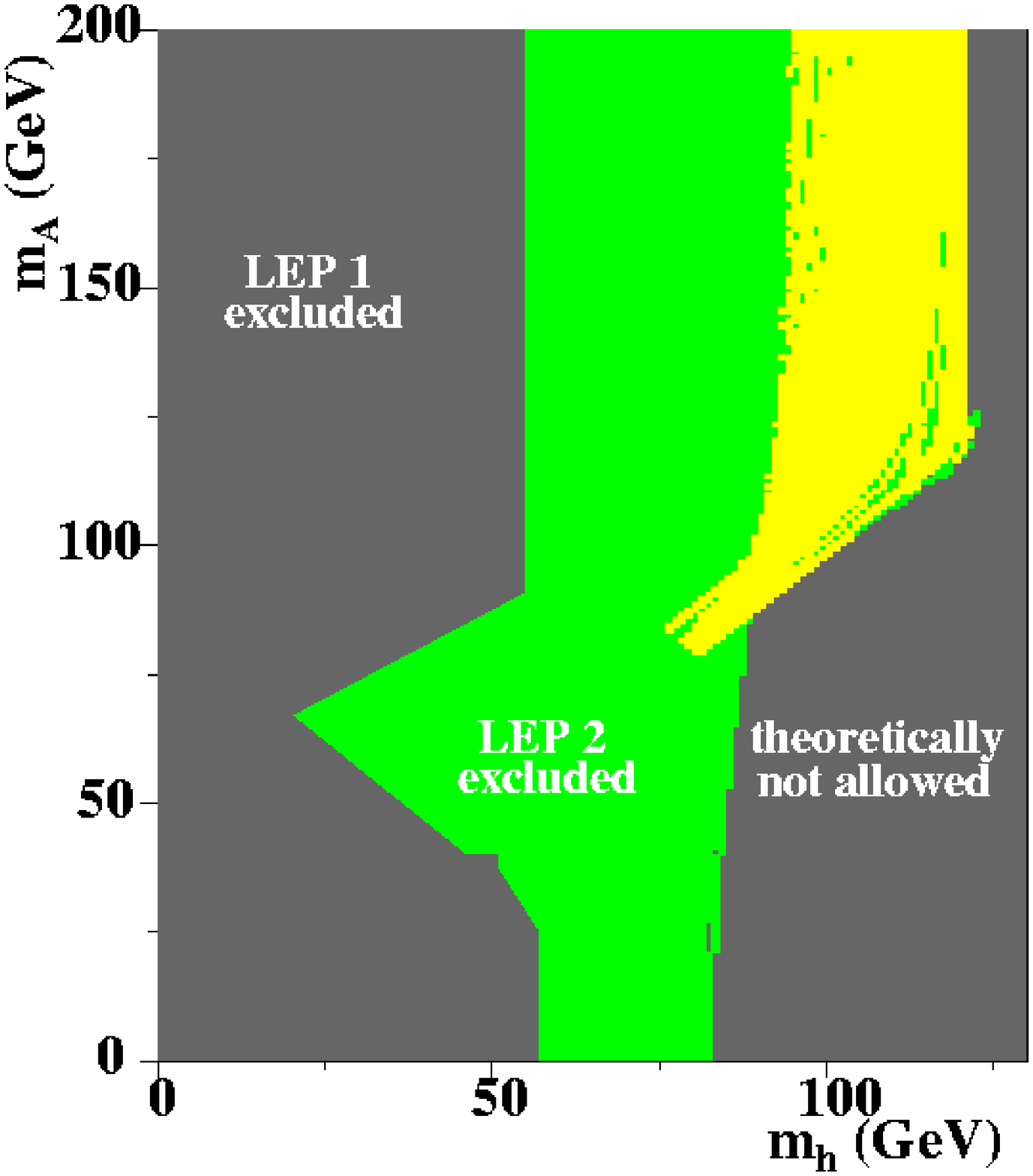}
\includegraphics[width=0.49\textwidth]{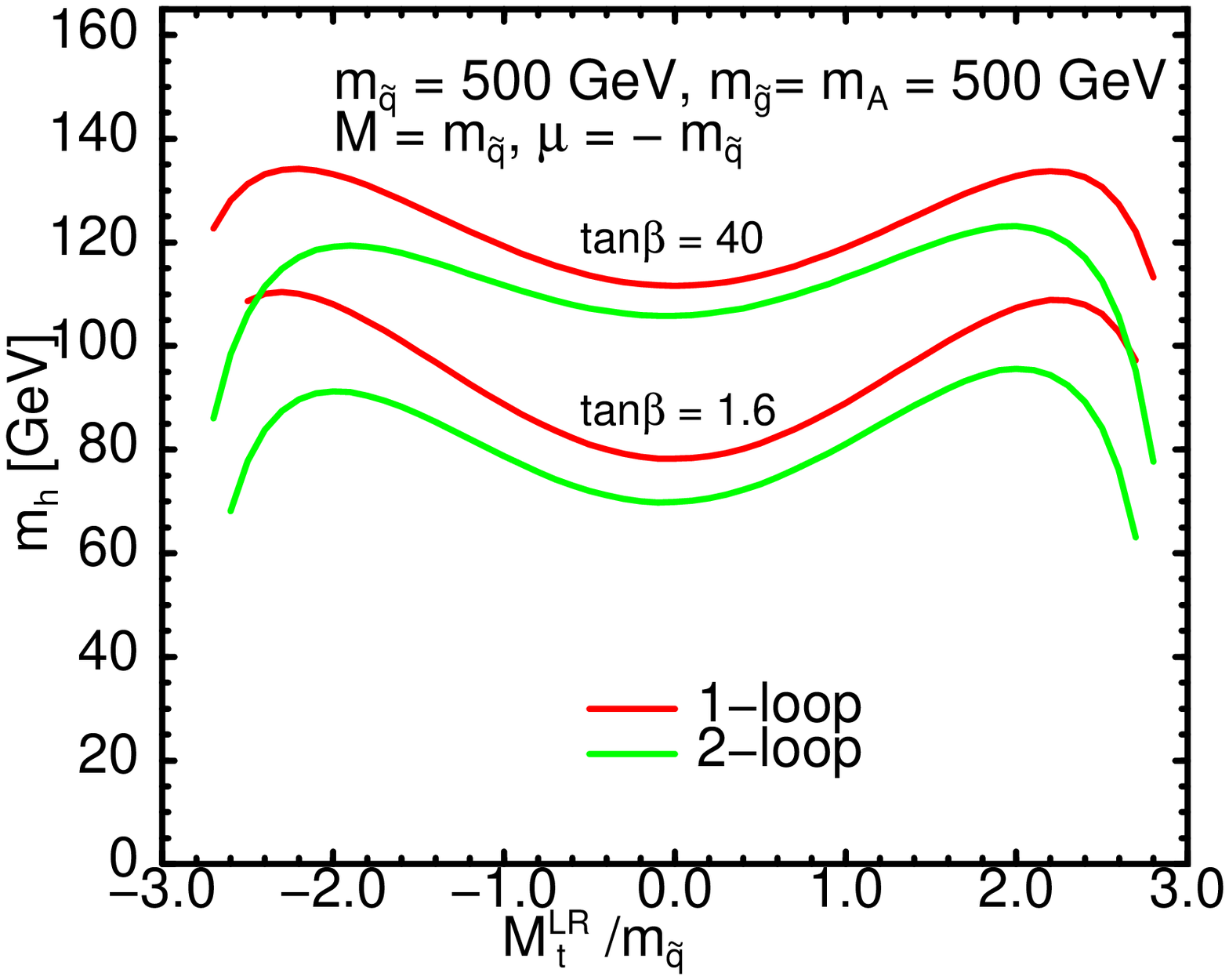}
\end{center}
\vspace*{-1.2cm}
\end{figure}

\clearpage
General parameter scans in the MSSM are very important 
in order to set reliable limits on h and A masses.
For LEP1 newly unexcluded mass regions were pointed 
out~\cite{as1}, and for 172 GeV LEP2 data stringent mass
limits disappeared completely~\cite{as2} allowing a Higgs
boson to escape detection although it could have been
kinematically produced.
The latest general scan results with 189 GeV DELPHI
data are shown in Fig.~\ref{fig:delphi189scan}~\cite{tamperedelphi}.
The corresponding parameter ranges are listed in Table~\ref{tab:range}.
The limit on the h mass is at 75 GeV compared to 82 GeV for the DELPHI 
benchmark result~\cite{delphihiggs}. 
The h and A mass limits are
given by the light grey edge intruding the LEP2 excluded area.
An extension of the $A$ range to $\pm2$ does not change the mass limit,
but it increases the maximal Higgs mass to 130 GeV.
Reduced mass limits are also presented by OPAL~\cite{opal}.

In addition to the scanning over the Supersymmetric parameters, more precise
mass, cross section and branching ratio calculation based on 2-loop
Full-Diagrammatic Calculations (FDC)~\cite{hollik} are used.
Figure~\ref{fig:delphi189scan} shows also that the maximal 
h mass is reduced by about 10 GeV and the maximum is already
reached for smaller mixing values in the scalar top sector.
The reduction of the maximal Higgs boson mass is important
for LEP physics, since it affects the excluded $\tan\beta$ range
(see Fig.~\ref{fig:mhtb}) and reduces the uncovered parameter 
region. LEP comes closer to a definitive test of the MSSM.

\begin{table}[bp]
\caption{\label{tab:range} General scan parameter ranges.}
\begin{tabular}{c|c|c|c|c|c}
Parameter& $m_{\mathrm{sq}}$ (GeV) & $m_{\mathrm{g}}$  (GeV) &
$\mu$   (GeV) & $A$ & $\tan\beta$ \\
\hline
Range    & 200\,---\,1000   &  200\,---\,1000   & $-500$\,---\,500 &
$-1$\,---\,$+1$ & 0.5\,---\,50
\end{tabular}
\vspace*{-6mm}
\end{table}

\clearpage
\section{Charged Higgs Bosons}
\begin{figure}[hp]
\caption{\label{fig:xsechphm} $\rm e^+e^-\rightarrow H^+H^-$ 
         cross section 
         including 1-loop radiative corrections.}
\begin{center}
\vspace*{-0.2cm}
\includegraphics[width=0.49\textwidth]{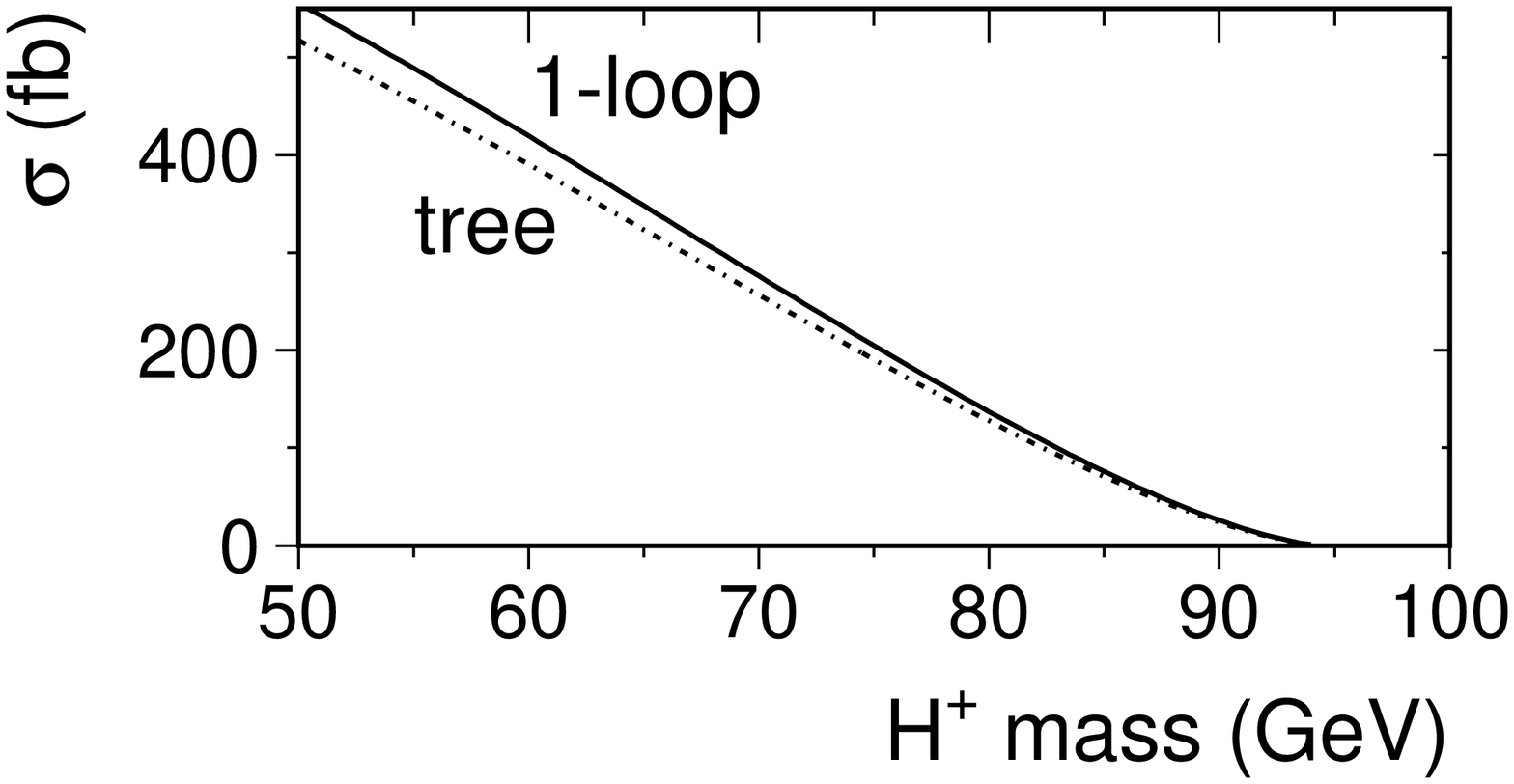}
\includegraphics[width=0.49\textwidth]{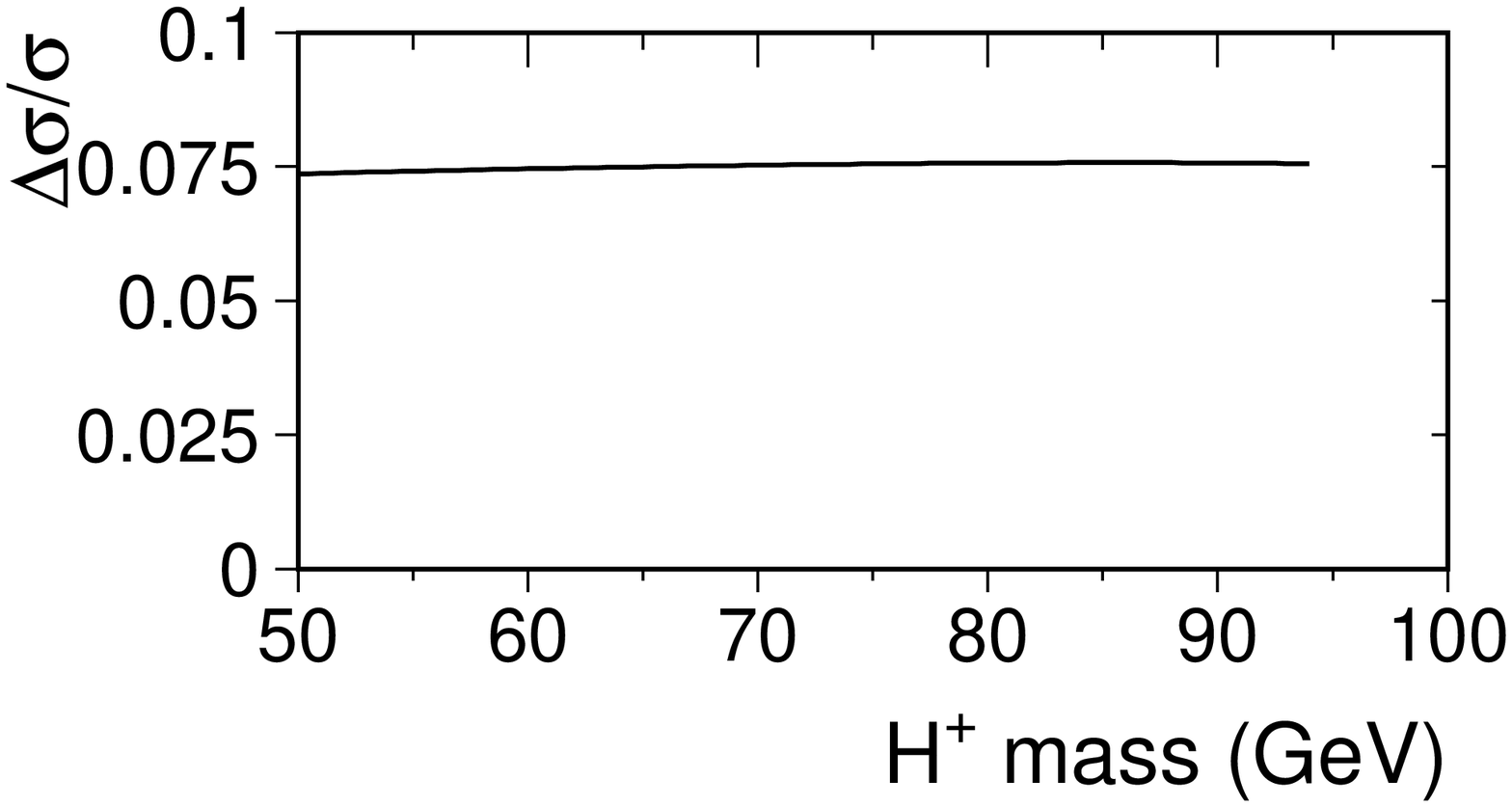}
\end{center}
\vspace*{-1cm}
\end{figure}

\begin{figure}[hp]
\caption{\label{fig:hphm} Reconstructed $\rm H^\pm$ mass
in the hadronic and semi-leptonic search channels (left), and 
resulting mass limits (right) including the leptonic channel and
the combination of all three channels. In addition, the expected
mass limit is shown.}
\begin{center}
\vspace*{-0.3cm}
\mbox{\epsfig{file=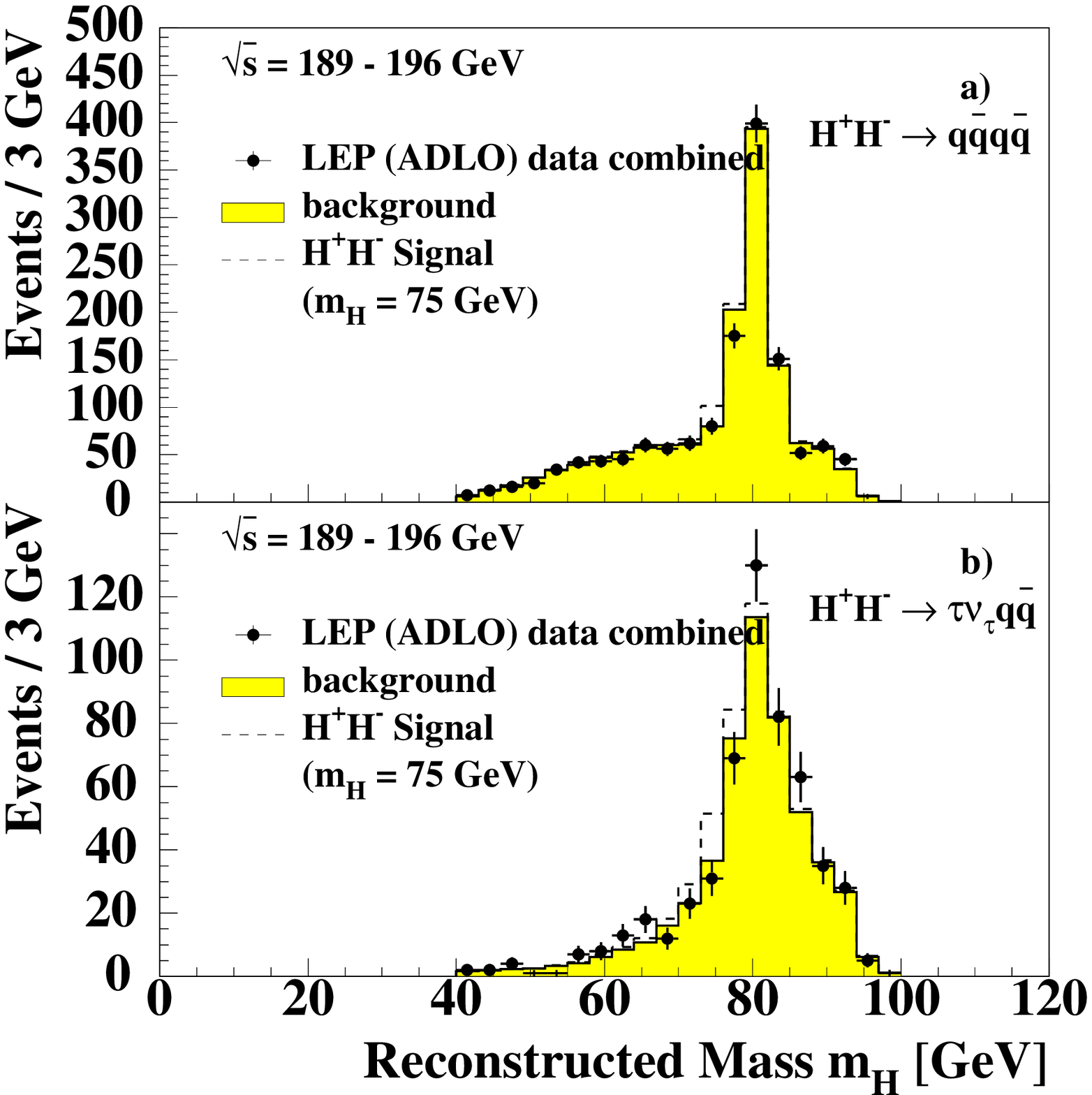,width=0.49\textwidth}}
\mbox{\epsfig{file=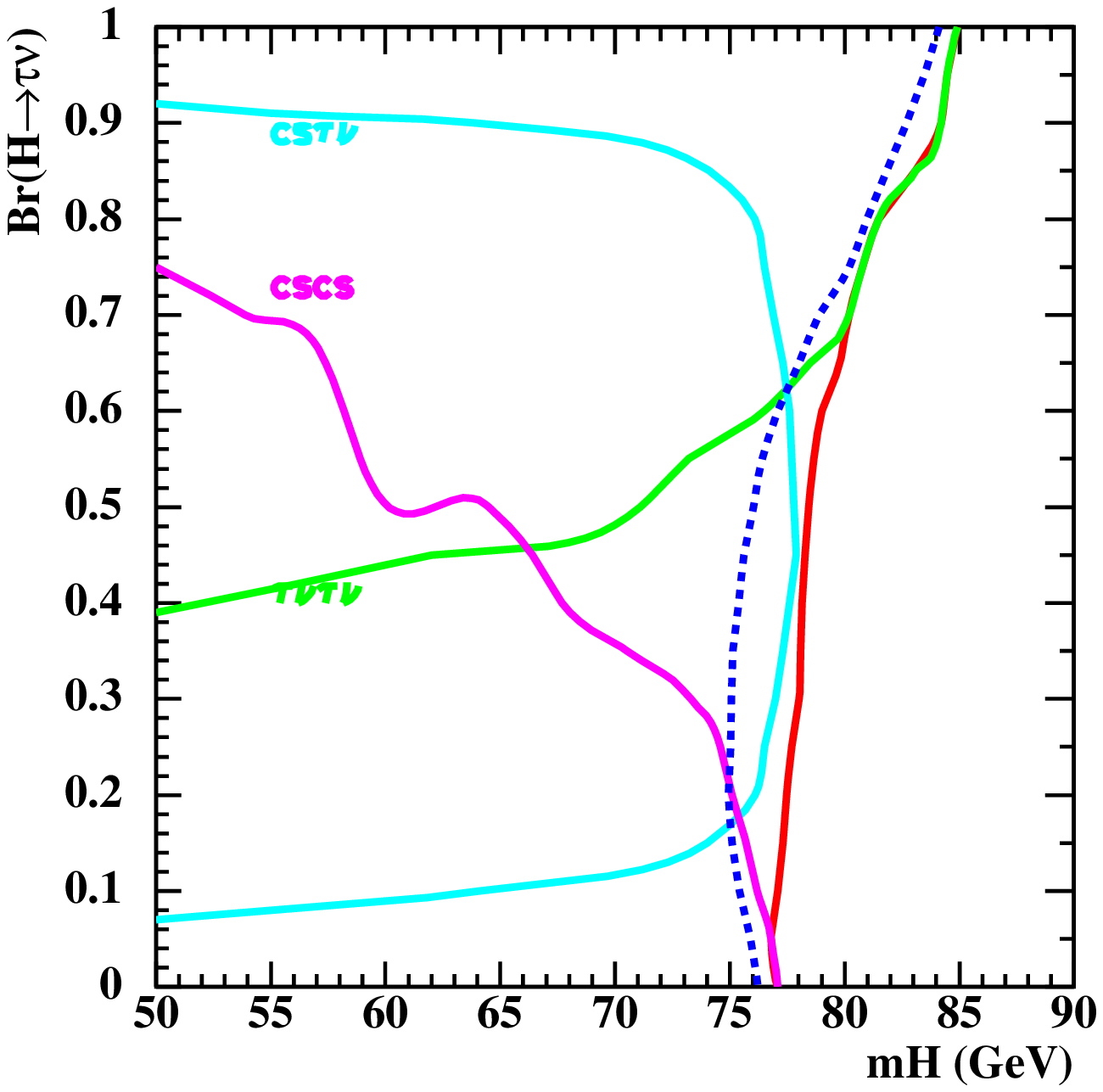,width=0.49\textwidth}}
\end{center}
\vspace*{-1cm}
\end{figure}

New calculations of the production cross section
$\rm e^+e^- \rightarrow H^+H^-$ include 1-loop
radiative corrections~\cite{arndt}. 
The cross section for $\sqrt{s}=190$ GeV 
is shown in Fig.~\ref{fig:xsechphm}.
Compared to the tree-level calculations, the cross section
is about 7.5\% larger for $2<\tan\beta<15$ independent 
of the charged Higgs boson mass.
Note that a 15\% reduction is obtained for very small and 
very large $\tan\beta$ values.
The results from the LEP experiments are still based
on tree-level calculations. Figure~\ref{fig:hphm} shows the 
reconstructed $\rm H^\pm$ mass for data, simulated signal
and background. Mass limits based on a statistical analysis
are also shown in Fig.~\ref{fig:hphm}.
The sensitivity in the leptonic search channel is larger, since
the irreducible WW background leads only in 1\% of cases 
to the $\tau\nu\tau\nu$ final state.
Independent of the charged Higgs boson branching fraction,
a mass limit of 77 GeV is set at 95\% CL.

\clearpage
\section{Supersymmetric Particles}

The clean environment of the LEP $\rm e^+e^-$ collider
and the good hermiticity of the LEP detectors allow 
detailed searches for Supersymmetric particles. Since the
lightest Supersymmetric particle is expected to be neutral
and stable, it escapes detection, leading to a missing energy
signature.  A typical candidate event taken at 200 GeV center-of-mass
energy with the DELPHI detector is shown in Fig.~\ref{fig:chi1chi2}.
The event shows two acoplanar jets indicating that energy, possibly
from neutralinos, has escaped detection. A variety of searches for
events with leptons, quarks, and missing energy is performed.
No indication of Supersymmetry has been found yet. Many mass limits close 
to the kinematic production threshold are set
with 189 GeV data~\cite{tampere7-716}.

\begin{figure}[hp]
\caption{\label{fig:chi1chi2} DELPHI candidate event 
$\rm e^+e^- \rightarrow \chi^0_1\chi^0_2\rightarrow 
\chi^0_1\chi^0_1qq$, compatible with background for 
example from $\rm ZZ\rightarrow qq\nu\nu$ production.}
\begin{center}
\mbox{\epsfig{file=t7807_002_r104722_e000661.epsi,width=\textwidth}}
\end{center}
\vspace*{-1.5cm}
\end{figure}

\begin{figure}[hp]
\caption{\label{fig:slep} Comparison of the number of 
data (left) and simulated background (right) events in the slepton search.}
\begin{center}
\vspace*{-0.3cm}
\mbox{\epsfig{file=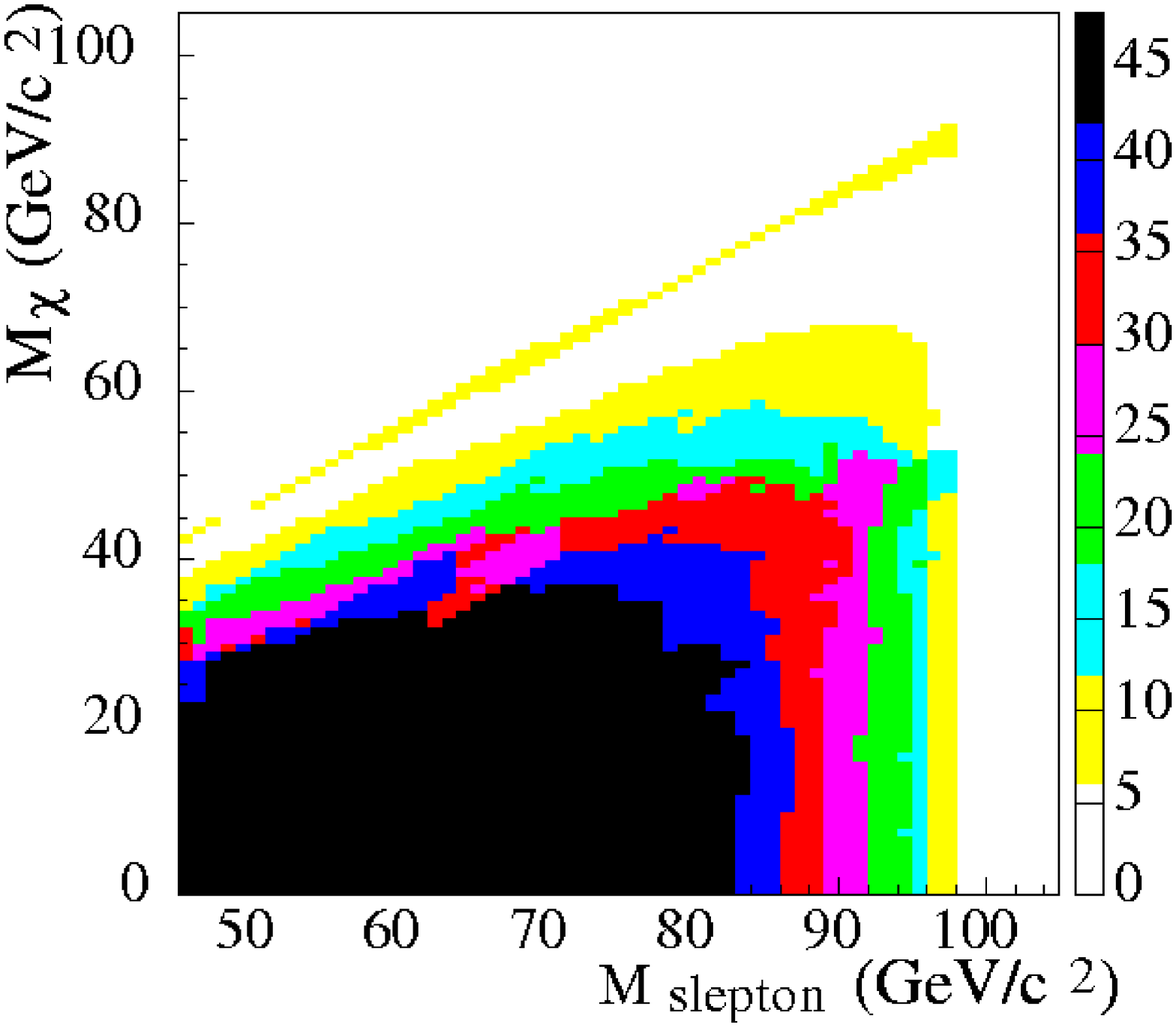,width=0.49\textwidth}}
\mbox{\epsfig{file=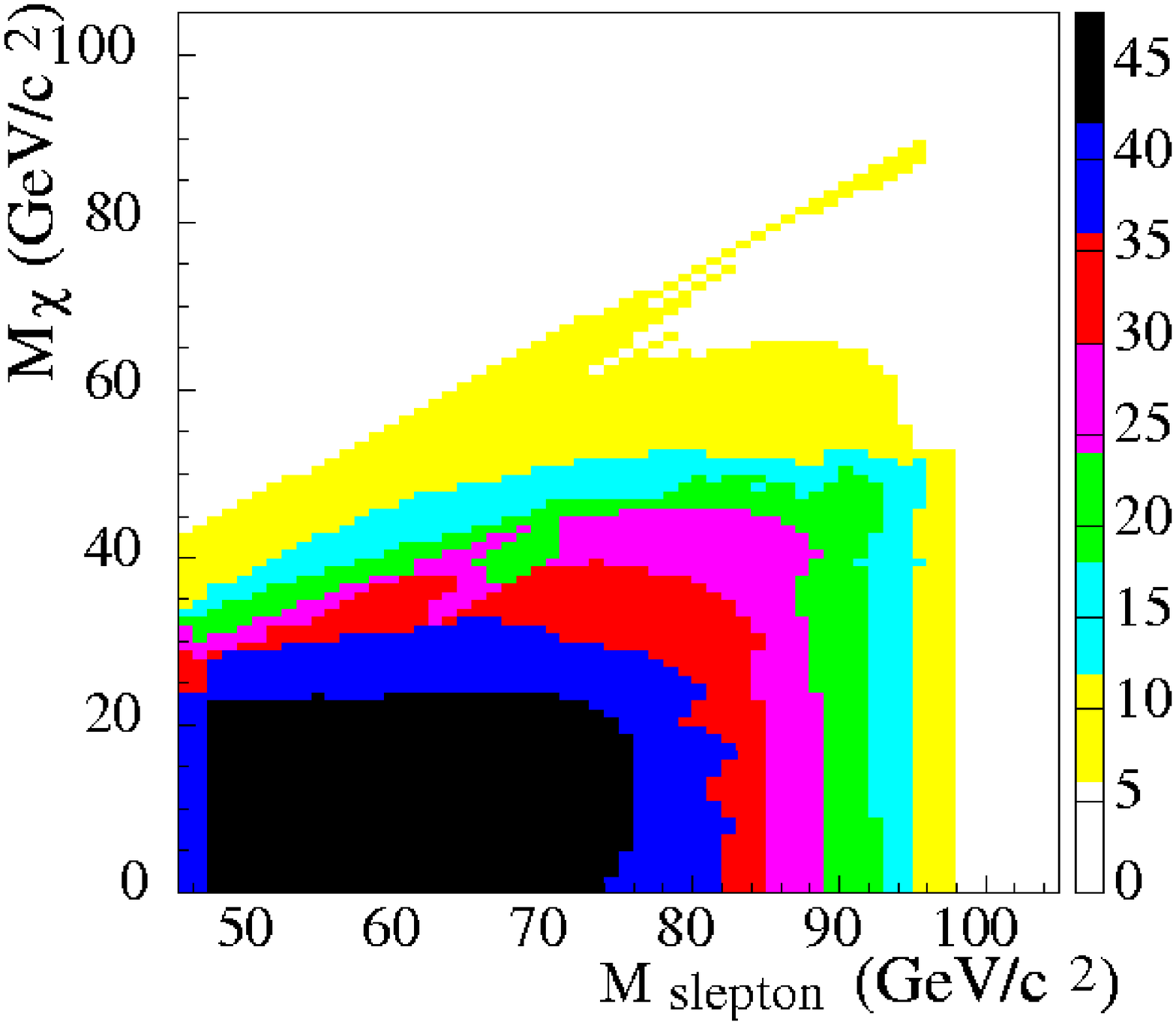,width=0.49\textwidth}}
\end{center}
\vspace*{-0.5cm}
\end{figure}

\begin{figure}[hp]
\vspace*{-2mm}
\caption{\label{fig:stop} Comparison of the number of 
data (upper) and simulated background (lower) events 
in the stop (left) and sbottom (right) search.}
\begin{center}
\vspace*{-0.4cm}
\mbox{\epsfig{file=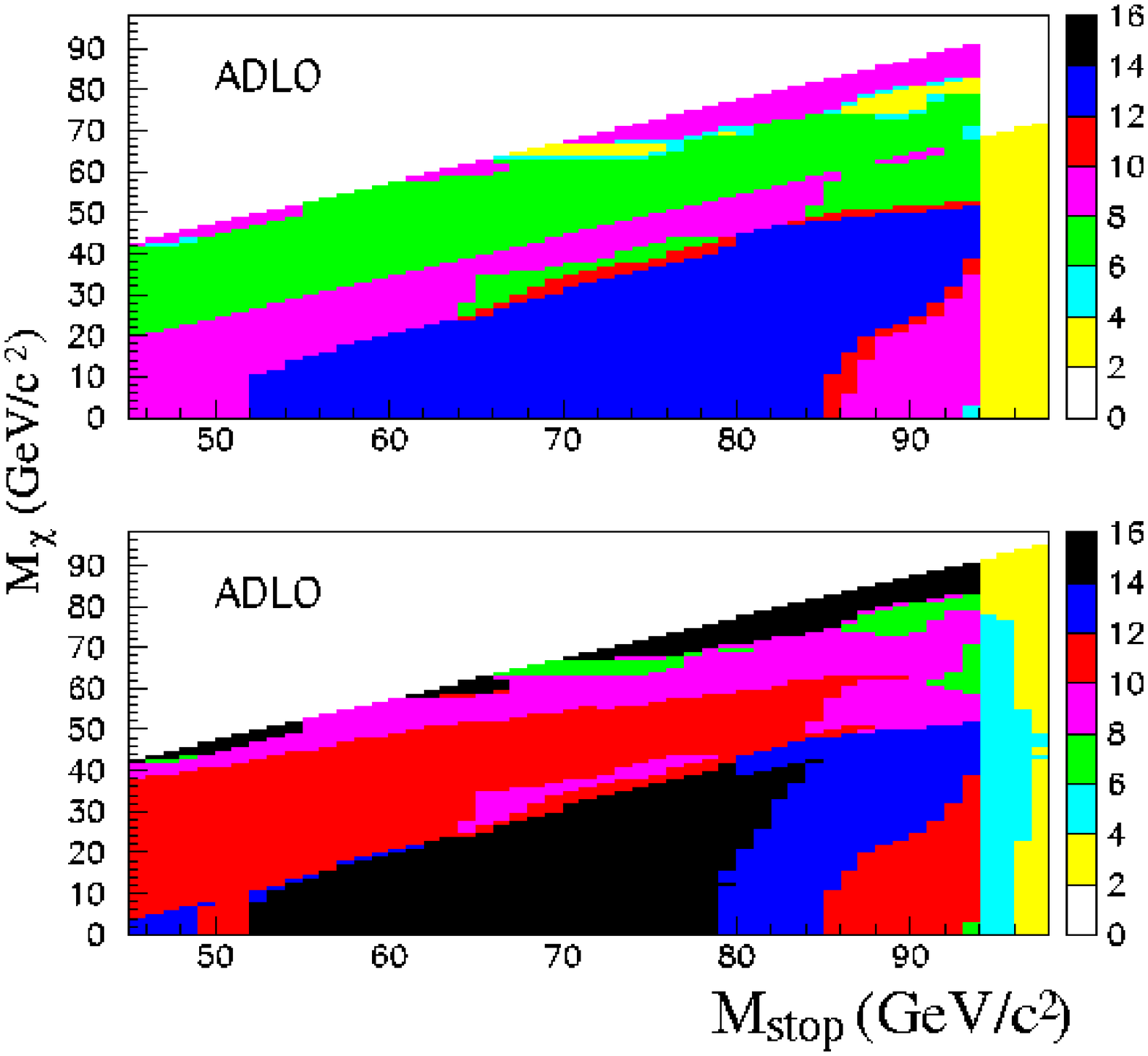,width=0.51\textwidth}}
\mbox{\epsfig{file=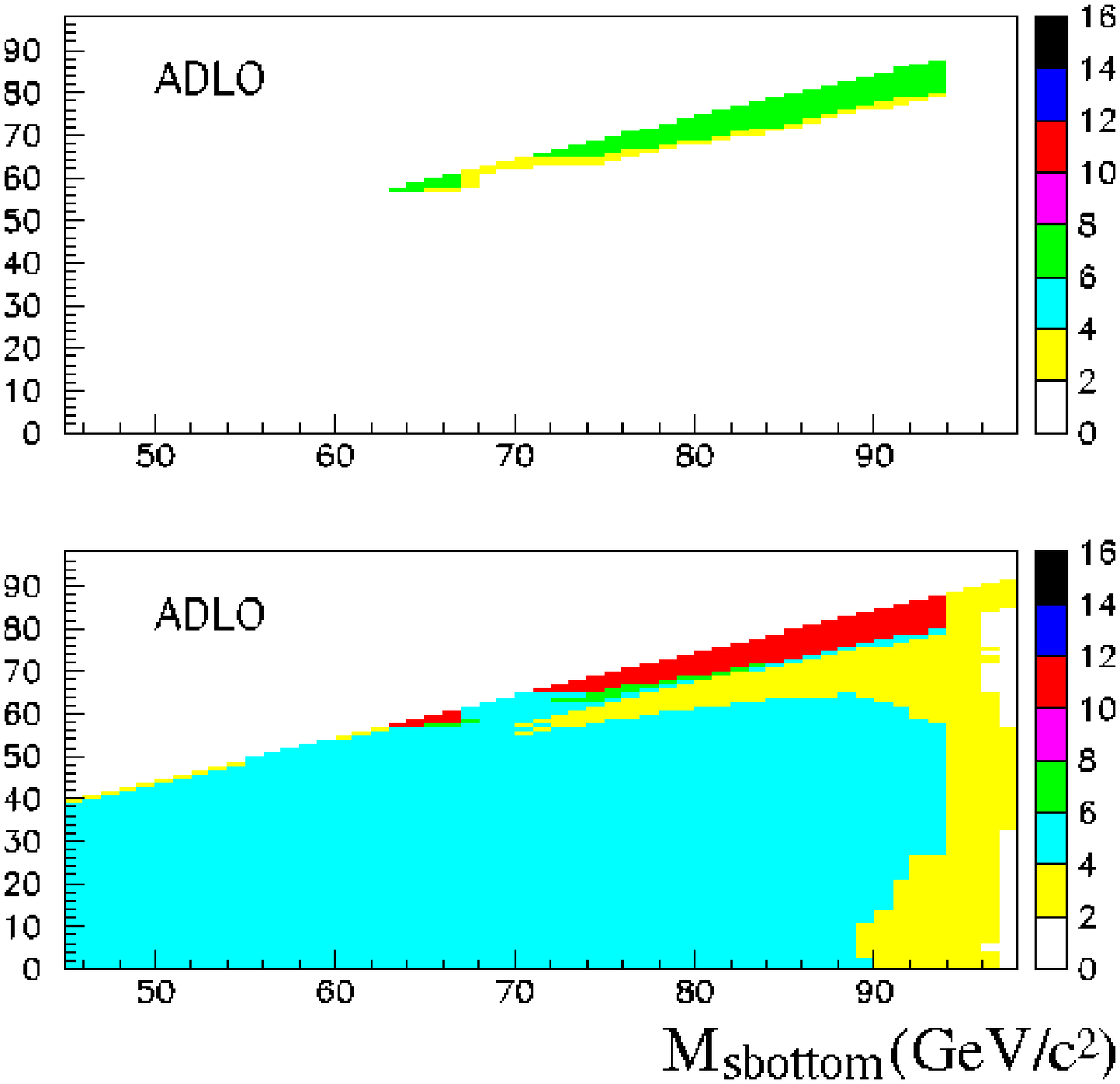,width=0.48\textwidth}}
\end{center}
\vspace*{-0.8cm}
\end{figure}

\begin{figure}[hp]
\caption{\label{fig:gamma} Recoil mass spectra of single (left) 
                           and double (right) photon production.}
\begin{center}
\vspace*{-0.6cm}
\mbox{\epsfig{file=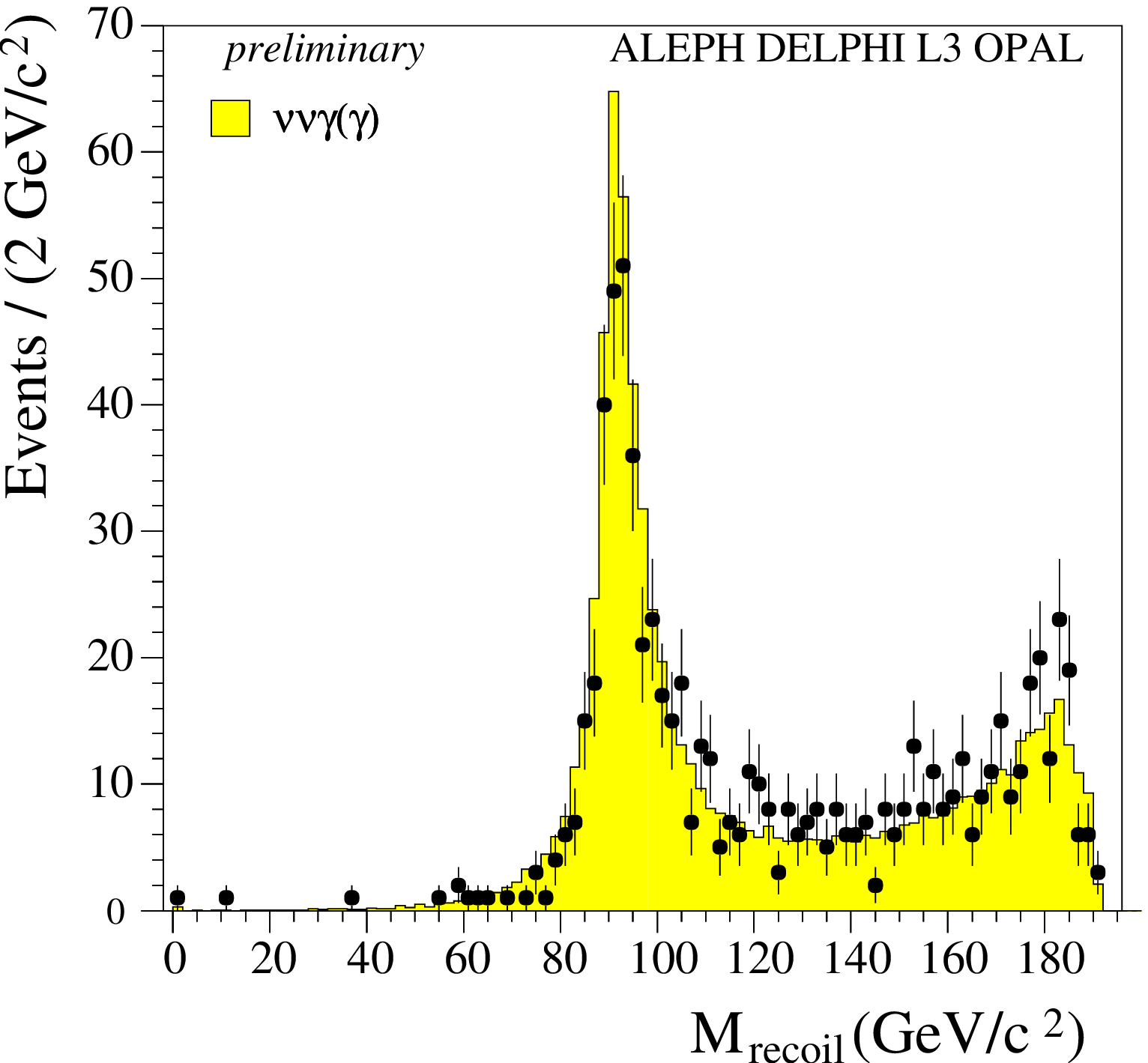,width=0.49\textwidth}}
\mbox{\epsfig{file=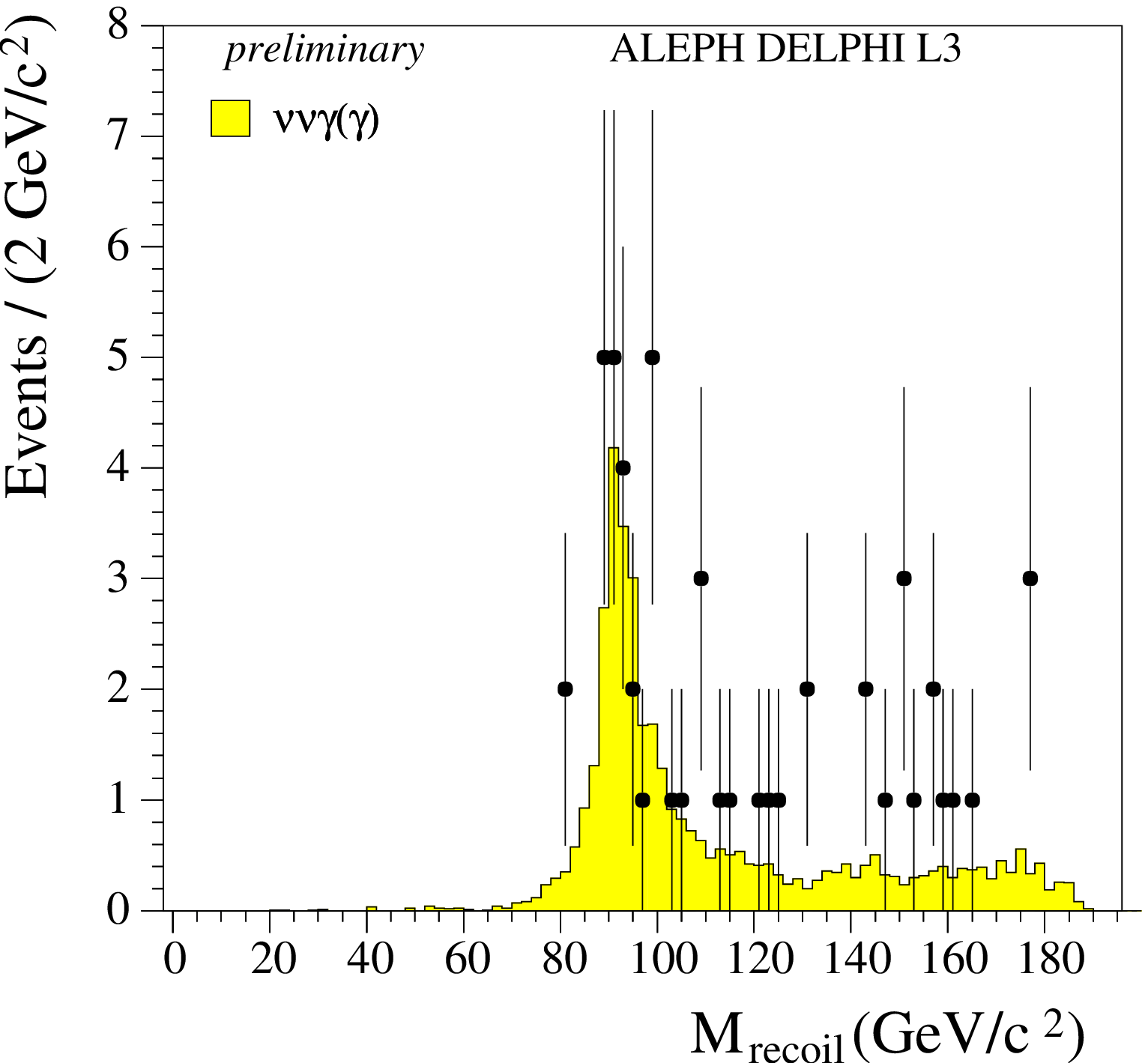,width=0.49\textwidth}}
\end{center}
\vspace*{-1.1cm}
\end{figure}

\clearpage
A few examples of data and background comparisons
for the data-taking up to 196 GeV are presented.
The number of observed and expected events in the 
search for sleptons is shown in Fig.~\ref{fig:slep} as a function
of slepton and neutralino masses.
The large value of the top mass could imply a large mixing
for scalar top quarks, thus rendering the lighter scalar top
to be the lightest Supersymmetric quark. A comparison of
data and simulated background events in the search for
scalar top and scalar bottom quarks is given
in Fig.~\ref{fig:stop}. 
The number of data events is compatible with the simulated
background expectation.
The decay $\chi^0_2\rightarrow \chi^0_1\gamma$
is also possible. A signal could be observed in the single or double
photon spectra. 
Figure~\ref{fig:gamma} shows no indication of deviation between data
and background simulation.

\clearpage

\section{Conclusions}
\vspace*{-1mm}
No indication of new particles is observed.
The combination of the data from the four LEP experiments
provides significantly more detection sensitivity.
The SM Higgs mass is larger than 102.6 GeV at 95\% CL.
In the remaining year of LEP operation, the SM Higgs boson
could be discovered up to 110 GeV at $5\sigma$ or excluded up to
about 114 GeV at 95\% CL.
Stringent mass limits are set in general extensions of the SM,
even for strongly reduced
HZ production rates. The\,charged\,Higgs\,boson\,mass is\,larger\,than\,77\,GeV 
which\,is\,close to\,the\,W\,mass, where WW production gives\,a\,large 
irreducible background.
Large regions of the MSSM parameter space are excluded, in particular,
small $\tan\beta$ values.
In future studies, 
the model-independent presentations of the experimental results
should be even more emphasized and general parameter scans
should be favored over benchmark interpretations.

%
%
%

\end{document}